\documentclass[a4paper, 12pt]{article}
\usepackage{jheppub}
\usepackage{amssymb} 
\usepackage{amsmath}
\usepackage{mathtools}
\usepackage{amsfonts}    
\usepackage{dsfont}
\usepackage{young}
\usepackage[vcentermath]{youngtab}
\usepackage[mathscr]{euscript}
\usepackage{tensor}
\usepackage{xcolor}

\newcommand{\be}{ \begin{equation}}
\newcommand{\ee}{\end{equation}}
\newcommand{\bea}[1]{\begin{eqnarray}\label{#1} }
\newcommand{\eea}{\end{eqnarray}}

\def\ZZZ{{\hskip-3pt\hbox{ Z\kern-1.6mm Z}}}

\def\zzz{{\hskip-3pt\hbox{ z\kern-1mm z}}}

\newcommand{\cK}{{\cal K}}
\newcommand{\cJ}{{\cal J}}

\def\one{{\hbox{ 1\kern-.8mm l}}}
\def\zero{{\hbox{ 0\kern-1.5mm 0}}}

\allowdisplaybreaks

\title{Tensionless strings on $\boldsymbol{\text{AdS}_3 \times \text{S}^3 \times \text{S}^3 \times \text{S}^1}$}

\author[a,b]{Matthias R.\ Gaberdiel}
\author[a]{and Vit Sriprachyakul}
\affiliation[a]{Institut f\"ur Theoretische Physik,
ETH Z\"urich,\\
Wolfgang-Pauli-Strasse 27,
8093 Z\"urich, Switzerland}
\affiliation[b]{Kavli Institute for Theoretical Sciences, University of Chinese Academy of Sciences,\\
Beijing 100190, China}
\emailAdd{gaberdiel@itp.phys.ethz.ch}
\emailAdd{vsriprachyak@phys.ethz.ch}

\abstract{We propose that string theory on ${\rm AdS}_3 \times {\rm S}^3 \times {\rm S}^3 \times {\rm S}^1$ with \emph{two} units of NS-NS flux through each of the two $3$-spheres is exactly dual to the symmetric orbifold of $2$  bosons and $8$ free fermions. Here, one of the two bosons describes the compact ${\rm S}^1$, while the other one corresponds to the radial (non-compact) direction of ${\rm AdS}_3$. Unlike the analogous situation for ${\rm AdS}_3 \times {\rm S}^3 \times \mathbb{T}^4$, the description makes sense both in the NS-R as well as the hybrid formulation, and we explain the duality in both frameworks.}

\begin{document}

\maketitle

\section{Introduction}

A few years ago it was understood that string theory on ${\rm AdS}_3\times {\rm S}^3 \times \mathbb{T}^4$ with one unit of NS-NS flux through both ${\rm AdS}_3$ and ${\rm S}^3$ is exactly dual to the symmetric orbifold of $\mathbb{T}^4$ \cite{Gaberdiel:2018rqv,Eberhardt:2018ouy,Eberhardt:2019ywk}. The string background could be described in terms of a (solvable) WZW model following \cite{Maldacena:2000hw,Berkovits:1999im}, and it gives rise to the single string spectrum of the symmetric orbifold \cite{Gaberdiel:2018rqv,Eberhardt:2018ouy}. Furthermore, the correlation functions of the symmetric orbifold could be reproduced from those of the worldsheet \cite{Eberhardt:2019ywk,Dei:2020zui}, see also \cite{Eberhardt:2020akk,Knighton:2020kuh,McStay:2023thk,Dei:2023ivl,Knighton:2023mhq} for further developments. 

Apart from replacing $\mathbb{T}^4$ by K3, there is another maximally supersymmetric background involving strings on ${\rm AdS}_3\times {\rm S}^3$, namely 
\be
{\rm AdS}_3 \times {\rm S}^3 \times {\rm S}^3 \times {\rm S}^1 \ .
\ee
The search for the holographic dual of it has a long and complicated history, see \cite{Elitzur:1998mm,deBoer:1999gea,Gukov:2004ym,Tong:2014yna,Eberhardt:2017fsi,Eberhardt:2017pty}. By analogy with the   $\mathbb{T}^4$ case one may suspect that the dual CFT lies on the moduli space that contains the symmetric orbifold of the ${\cal S}_\kappa$ theory \cite{Elitzur:1998mm}\footnote{Alternatively, the CFT dual should be identified with the moduli space of instantons on ${\rm S}^3\times {\rm S}^1$, and its large ${\cal N}=4$ structure was recently carefully studied in \cite{Witten:2024yod}.} --- the ${\cal S}_\kappa$ theory effectively describes the CFT with target space ${\rm S}^3 \times {\rm S}^1$ --- but this seemed to be in conflict with the supergravity BPS spectrum that was derived in \cite{deBoer:1999gea}, see in particular \cite{Gukov:2004ym} for a careful discussion. It was later realised in \cite{Eberhardt:2017fsi} that there are actually much fewer BPS states in supergravity, and this removed some of the objections of \cite{Gukov:2004ym}, and in particular, reinstated the proposal of \cite{Elitzur:1998mm} at least for the case where the size of one of the two ${\rm S}^3$ is minimal \cite{Eberhardt:2017pty}. One may thus expect that there could be an exact duality relating the background with the corresponding NS-NS fluxes to the symmetric orbifold theory itself. Since one of the $\mathfrak{su}(2)$ levels is equal to $k=1$, the NS-R description of the background has the usual problems, but a hybrid description similar to the torus case \cite{Berkovits:1999im} was found in  \cite{Eberhardt:2019niq}, and it exactly reproduced the spectrum of the symmetric orbifold. However, obtaining the correlators of the symmetric orbifold of ${\cal S}_\kappa$ from the worldsheet description turned out to be more difficult, and no convincing argument has so far been found --- in particular, the arguments of \cite{Dei:2020zui} do not apply, even if one just considers the simplest example of \cite{Eberhardt:2019niq} (where both $\mathfrak{su}(2)$ levels are equal to $k_\pm =1$) that possesses a free field realisation.  It is therefore not clear whether this example gives in fact rise to an exact duality, or whether the worldsheet theory is only dual to a deformation of the symmetric orbifold by a suitable exactly marginal operator, similar to what happens in \cite{Eberhardt:2021vsx}.

In a somewhat different development,  the near-boundary limit of string theory on ${\rm AdS}_3$ was studied in \cite{Knighton:2024qxd,Sriprachyakul:2024gyl,Sriprachyakul:2024xih}, see also \cite{Hikida:2023jyc,Knighton:2024pqh} for related work.\footnote{Some aspects of the underlying path-integral methods were developed in \cite{Dei:2023ivl,Knighton:2023mhq}. Earlier attempts at identifying the CFT dual for general classes of AdS$_3$  backgrounds include \cite{Giveon:1998ns,Giveon:1999jg,Argurio:2000tb}, see also \cite{Halder:2022ykw}.} In particular, generalising the bosonic analysis of \cite{Eberhardt:2021vsx}, it was argued in \cite{Sriprachyakul:2024gyl,Sriprachyakul:2024xih} that superstrings on ${\rm AdS_3}\times X$ should be dual to the conformal field theory 
\begin{equation}
{\rm Sym}^N\left(\mathbb{R}^{(1)}_{\tilde Q}\times X\right)+\int\sigma_2\,,
\end{equation}
where $\mathbb{R}^{(1)}_{\tilde Q}$ is an $\mathcal{N}=1$ linear dilaton theory with dilaton slope $\tilde Q$, while $\sigma_2$ denotes a perturbing field from the twist-two sector. Applying this idea to $X={\rm S^3\times S^3\times S^1}$ then suggests that the dual CFT  should be 
\begin{equation}\label{1.2}
{\rm Sym}^N\left(\mathbb{R}^{(1)}_{\tilde Q}\times {\rm S^3\times S^3\times S^1}\right)+\int\sigma_2\ . 
\end{equation}
The background charge of the linear dilaton in (\ref{1.2}) vanishes provided that 
the size of the ${\rm AdS}_3$ is the same as for the above $\mathbb{T}^4$ case, i.e.\ provided that the (supersymmetric) level $k$ of the worldsheet $\mathfrak{sl}(2,\mathds{R})$ algebra is $k=1$. Given the relation of $k$ to the (supersymmetric) levels $k_\pm$ of the two $\mathfrak{su}(2)$ factors (describing the two ${\rm S}^3$), see \cite{Elitzur:1998mm}
\be\label{critical}
\frac{1}{k} = \frac{1}{k_+} + \frac{1}{k_-} \ , 
\ee
$k=1$ is only possible for $k_+ = k_-=2$.  But the supersymmetric $\mathfrak{su}(2)_2$ theory is actually equivalent to $3$ free fermions, so this suggests that the CFT dual for $k_\pm =2$ equals 
\begin{equation}\label{1.5}
{\rm Sym}^N\left(\mathbb{R}_{0}\times {\rm S^1}\times\text{8 fermions}\right) +\int\sigma_2 \ , 
\end{equation}
where the free fermions come from 
\be
1(\mathbb{R}^{(1)}_{\tilde Q}) + 3({\rm S}^3) + 3({\rm S}^3) + 1({\rm S}^1) = 8 \ , 
\ee
and the remaining ${\rm S}^1$ factor in the seed theory is a compact boson. At level $k=1$, it is natural to restrict the $\mathfrak{sl}(2,\mathds{R})$ worldsheet representations to the continuous representations $j=\frac{1}{2} + is$ --- in particular, this by itself already defines a modular invariant worldsheet theory --- and as we will show below, this then reproduces on the nose the spectrum of the symmetric orbifold, 
\be\label{1.7}
\boxed{{\rm AdS}_3 \times {\rm S}_3 \times {\rm S}^3 \times {\rm S}^1 \ \ \hbox{with $k_\pm=2$} \ \ \longleftrightarrow \ \ 
{\rm Sym}^N\left(\mathbb{R}_{0}\times {\rm S^1}\times\text{8 fermions}\right) \ . }
\ee
Note that the seed theory on the right is essentially equal to ${\cal S}_0^2$, and that eq.~(\ref{1.7}) is therefore very close to what was originally proposed in \cite{Gaberdiel:2018rqv}, see also \cite{Giribet:2018ada} --- the main difference is that now the boson from $\mathbb{R}_0$ is non-compact, and the worldsheet spectrum includes all continuous representations. Here the $s$ parameter of the continuous representations $j=\frac{1}{2} + is$ becomes the momentum of the uncompactified boson in the dual CFT.\footnote{Since on the worldsheet the representations with $s$ and $-s$ are equivalent, this suggests that the $\mathbb{R}_0$ factor is not simply a non-compact free boson, but rather something like the $c=1$ limit of Liouville theory \cite{Schomerus:2003vv,Runkel:2001ng}, see also \cite{McElgin:2007ak,Harlow:2011ny} and Section~\ref{sec: RNS analysis} below. We thank Lorenz Eberhardt for a discussion about this point.}

Obviously, this by itself does not yet prove that the dual CFT is indeed just the symmetric orbifold without any deformation, but there are a number of arguments that make this seem rather plausible. First of all, the BPS spectra of the two descriptions match precisely (since the entire spectrum matches). This is, to our knowledge, the only example where an  NS-R worldsheet theory reproduces the full BPS spectrum of the spacetime CFT --- usually there are `gaps' in the worldsheet theory, see e.g.\ the analysis in \cite[Sections~4.3 \& 4.4]{Eberhardt:2017pty}. Note that, contrary to what was claimed in \cite[Section~3.5]{Gaberdiel:2018rqv}, this BPS spectrum in fact contains the full supergravity spectrum of 
\cite{Eberhardt:2017fsi}, see the discussion in Section~\ref{sec:BPS} below. We also show there that the symmetric orbifold does not have \emph{any} exactly marginal operator in any of the twisted sectors --- exactly marginal large ${\cal N}=4$ operators are descendants of BPS states with $h=\frac{1}{2}$ and $j^\pm = \frac{1}{2}$, and no such operator exists in the twisted sectors of the symmetric orbifold of ${\cal S}_0^2$.\footnote{On the other hand, the symmetric orbifold of ${\cal S}_\kappa$ has such an exactly marginal operator in the twisted sector. For $\kappa\neq 0$ it comes from the $2$-cycle twisted sector, while for $\kappa=0$ it arises from the $3$-cycle twisted sector \cite{Gukov:2004ym,Eberhardt:2017fsi}. This argument therefore does not apply to the dualities considered in \cite{Eberhardt:2019niq}.}  As such, there is simply no deformation term one could add! The only exactly marginal operators of the symmetric orbifold come from the untwisted sector and they deform the seed theory, in particular, the radius of the ${\rm S}^1$.

Incidentally, it is also striking that in this example the $\mathfrak{sl}(2,\mathds{R})$ theory has again level $k=1$, just like for the ${\rm AdS}_3 \times{\rm S}^3 \times \mathbb{T}^4$ example. While it is not really possible to formulate the latter background in the NS-R formulation (since the supersymmetric $\mathfrak{su}(2)$ algebra also has level $k=1$), one may suspect that the localisation properties of the theory should essentially only depend on the $\mathfrak{sl}(2,\mathds{R})$ factor; this would therefore suggest that our example will also correspond to a symmetric orbifold (without any deformation). The fact that the $k=1$ $\mathfrak{sl}(2,\mathds{R})$ theory is quite special was already pointed out long ago \cite{Giveon:2005mi}, see also \cite{Eberhardt:2021vsx,Balthazar:2021xeh,Martinec:2021vpk} for related work. 

Finally, we find a hybrid description along the lines of  \cite{Eberhardt:2019niq}, and it also possesses a free field realisation. However, unlike the situation discussed in \cite{Eberhardt:2019niq}, it now involves four symplectic bosons, and thus the old localisation analysis of \cite{Dei:2020zui} applies directly. This effectively proves the desired localisation property in the hybrid description. Thus, as long as the NS-R description is indeed equivalent to the hybrid description, the same must be true in that formulation. 
\medskip

\medskip

Along the way we shall also discuss how spacetime supersymmetry of these backgrounds can be understood from a worldsheet perspective following \cite{Banks:1987cy}; this extends the analysis of \cite{Giveon:1998ns,Israel:2003ry,Ferko:2024uxi}. In particular, the relevant worldsheet structure sufficient to guarantee spacetime supersymmetry is in this case not a regular ${\cal N}=2$ algebra, but rather a certain extension of this algebra that we describe, see eq.~(\ref{2.12}) in Section~\ref{sec:worldsheetsusy}. 
\bigskip

The paper is organised as follows. Section~\ref{sec: RNS analysis} studies the above worldsheet theory in the NS-R formalism. In particular, we show how the spectrum of the dual CFT emerges, and we discuss the BPS spectum, both on the worldsheet and in the dual CFT. We also comment in Section~\ref{sec:worldsheetsusy} on how spacetime supersymmetry can be understood from the worldsheet perspective. In Section~\ref{hybrid} we turn to the hybrid description, following the analysis of \cite{Eberhardt:2019niq}. We find a new free field realisation of $\mathfrak{d}(2,1;\alpha)$ at $k_\pm =2$, and use it to analyse the spectrum of the theory, as well as the desired localisation property following \cite{Dei:2020zui}. Section~\ref{sec:conc} contains our conclusions and indicates directions for future work. Finally, there are a number of appendices where some of the more technical material is described. 

\section{The NS-R analysis}\label{sec: RNS analysis}

The NS-R analysis of the background in question actually follows quite closely what was done in \cite{Gaberdiel:2018rqv}, however, with one important difference. Using the formulation of \cite{Elitzur:1998mm,Maldacena:2000hw}, the worldsheet theory describing strings on ${\rm AdS}_3 \times {\rm S}^3 \times {\rm S}^3 \times {\rm S}^1$ with pure NS-NS flux is the WZW model associated to 
\be\label{WZW}
\mathfrak{sl}(2,\mathds{R})^{(1)}_k \oplus \mathfrak{su}(2)^{(1)}_{k_+} \oplus \mathfrak{su}(2)^{(1)}_{k_-}  \oplus \mathfrak{u}(1)^{(1)} \ , 
\ee
where the superscripts indicate that we consider the ${\cal N}=1$ superconformal affine algebras. (Our conventions for these algebras are described in some detail in Appendix~\ref{app:conventions}.) As reviewed in the Introduction, see the discussion around eq.~(\ref{critical}), we shall consider the case where 
\be\label{tensionless}
k_+ = k_- = 2 \ , \qquad k = 1\ .
\ee
After decoupling the fermions, see Appendix~\ref{app:conventions}, we obtain 
\be
\mathfrak{sl}(2,\mathds{R})_{k+2} \oplus \mathfrak{su}(2)_{k_+-2} \oplus \mathfrak{su}(2)_{k_- - 2}  \oplus \mathfrak{u}(1) \oplus \bigl( \hbox{10 free fermions} \bigr) \ .
\ee
For $k_\pm =2$, the two bosonic $\mathfrak{su}(2)$ factors have level $k=0$ after the decoupling and hence are trivial, i.e.\ just consist of the vacuum state. Thus the bosonic degrees of freedom of the worldsheet theory associated to (\ref{tensionless}) consist only of $\mathfrak{sl}(2,\mathds{R})_3 \oplus \mathfrak{u}(1)$. Imposing the physical state condition removes two bosonic and two fermionic degrees of freedom; following \cite{Sriprachyakul:2024gyl} it is natural to gauge away two bosonic degrees of freedom from the $\mathfrak{sl}(2,\mathds{R})_3 $ factor, and we thus end up with two bosons, one describing the `radial' direction of ${\rm AdS}_3$, while the other is the one associated to the ${\rm S}^1$ factor. In addition, we also retain $8$ free fermions. 

In \cite{Gaberdiel:2018rqv}, it was proposed that we should restrict the $\mathfrak{sl}(2,\mathds{R})_3$ representations to the `continuous' representations associated to $j=\frac{1}{2} + i s$ with $s=0$. However, this is not completely consistent as the resulting worldsheet theory would not be modular invariant. Indeed, since the `radial' boson of ${\rm AdS}_3$ survives the physical state condition, we need to include also its associated momentum. Thus we propose now to restrict the $\mathfrak{sl}(2,\mathds{R})_3$ representations to \emph{all} continuous representations with $j=\frac{1}{2} + i s$, without any restriction on $s\in\mathds{R}$, see also \cite{Giribet:2018ada}. This is to say, the worldsheet theory in question will have the form 
\be\label{wsspec}
{\cal H} =  \Bigl[ \sum_{w\in\mathbb{Z}} \int_0^1d \alpha \int_{\mathds{R}} ds\, \sigma^w \bigl({\cal H}_{j=\frac{1}{2}+is,\alpha} \bigr) \otimes \sigma^w \bigl(\bar{\cal H}_{j=\frac{1}{2}+is,\alpha} \bigr) \Bigr] \otimes {\cal H}^{S^1} \otimes{\cal H}_{\rm fermions} \ ,
\ee
where $w$ describes spectral flow, while the $\alpha$ parameter determines the eigenvalue of the $\mathfrak{sl}(2,\mathds{R})$ Cartan generator $J^3_0$ modulo integers.\footnote{The fact that $w$ and $\alpha$ are the same for left- and right-movers is as in the original proposal of \cite{Maldacena:2000hw}, see e.g.\ their Appendix B.4.}  The ghosts eliminate two bosons and two fermions, and the resulting worldsheet theory is then indeed modular invariant --- this is explained in more detail in Appendix~\ref{app:NS-Rmod}. (The ${\rm S}^1$ factor, as well as the eight fermions are modular invariant by themselves, so it is just a question of whether the above $\mathfrak{sl}(2,\mathds{R})$ part minus the two bosons is modular invariant.)

Obviously, this modification has an impact on the physical spectrum of the theory, and we therefore need to redo that part of the analysis in \cite{Gaberdiel:2018rqv}. In order to make the discussion self-contained, let us briefly review their argument. For the $w$-spectrally flowed $\mathfrak{sl}(2,\mathds{R})$ representation associated to spin $j$, the mass-shell condition in the NS-sector is 
\be\label{level}
- j (j-1) - w (m + \tfrac{w}{4}) + N + h^{S^1} = \tfrac{1}{2} \ , 
\ee
where $h^{S^1}$ is the conformal dimension associated to the state from the ${\rm S}^1$ factor, while $N$ is the excitation number of the remaining boson from $\mathfrak{sl}(2,\mathds{R})$ and the $8$ fermions. Finally, $m$ is the $J^3_0$ eigenvalue, i.e.\ the eigenvalue with respect to the Cartan generator of the $\mathfrak{sl}(2,\mathds{R})^{(1)}$ algebra, before spectral flow. There is also a similar relation for the right-movers.

The basic idea of \cite{Gaberdiel:2018rqv} is to solve (\ref{level}) for $m$ for any $j=\frac{1}{2} + i s$. Then, using the fact that the spacetime conformal dimension differs from the $J^3_0$ eigenvalue before spectral flow by $\frac{w}{2}$ --- the theory is at $k=1$ --- this leads to, see \cite[eq.~(2.15) and (2.16)]{Gaberdiel:2018rqv}
\be\label{NS}
h = \frac{N}{w} + \frac{h^{S^1} + s^2}{w} + \frac{w^2-1}{4w} \ ,  \qquad 
\bar{h} = \frac{\bar{N}}{w} + \frac{\bar{h}^{S^1} + s^2}{w} + \frac{w^2-1}{4w} \ .
\ee
Furthermore, the condition $m-\bar{m}\in\mathbb{Z}$ --- this comes from the fact that $\alpha$ is the same for the left- and the right-movers --- guarantees that the spacetime spectrum satisfies $h-\bar{h}\in\mathbb{Z}$ (for states coming from the worldsheet NS-NS sector). 

For the R sector one finds instead of (\ref{NS}), see \cite[eq.~(2.17) and (2.18)]{Gaberdiel:2018rqv},
\be
h = \frac{N}{w} + \frac{h^{S^1} + s^2}{w} + \frac{w^2+1}{4w} \ ,  \qquad 
\bar{h} = \frac{\bar{N}}{w} + \frac{\bar{h}^{S^1} + s^2}{w} + \frac{w^2+1}{4w} \ .
\ee
Again, the condition that $m-\bar{m}\in\mathbb{Z}$ guarantees that the spacetime spectrum satisfies $h-\bar{h}\in\mathbb{Z}$ for the states from the worldsheet R-R sector. On the other hand, for the states in the NS-R and R-NS sector we obtain $h-\bar{h}\in\mathbb{Z} + \frac{1}{2}$, since the R-sector $J^3_0$ eigenvalue is shifted by $\pm\frac{1}{2}$ relative to the NS-sector one, see e.g.\ the discussion in \cite[eq.~(2.6)]{Ferreira:2017pgt}. This obviously ties in with the idea that the NS-NS and R-R states describe spacetime bosons, while the NS-R and R-NS states are spacetime fermions. 

The only difference to the analysis of \cite{Gaberdiel:2018rqv} is therefore the fact that both $h$ and $\bar{h}$ contain the additional term $\frac{s^2}{w}$. The rest of the analysis then proceeds as in that paper, and we conclude that the states from the $w$-spectrally flowed sector give rise to the states in the $w$-cycle twisted sector of $8$ fermions and $2$ bosons. The boson associated to ${\rm S}^1$ comes with its own momentum terms, and they contribute to the conformal dimension in the $w$-cycle twisted sector via the term $h^{S^1}/w$. The additional $\frac{s^2}{w}$ term should then be interpreted as the momentum of the second boson, i.e.\ the one that is left over from the $\mathfrak{sl}(2,\mathds{R})$ factor after imposing the physical state condition. Finally, since $s$ is a continuous parameter (that is the same for both left- and right-movers), this boson is uncompactified. 

A cautionary comment is in order: since on the worldsheet, the representation with $j= \frac{1}{2} + is$ is equivalent to that with $j=\frac{1}{2} - is$, the bosonic degree that survives from the $\mathfrak{sl}(2,\mathds{R})$ factor is probably not simply a non-compact free boson, but rather something like the $c=1$ limit of Liouville theory \cite{Schomerus:2003vv,Runkel:2001ng,McElgin:2007ak,Harlow:2011ny}. On the level of the current analysis, it is difficult to distinguish precisely between the different possibilities, but this should eventually become clear once the correlation functions have been studied in more detail, see Section~\ref{sec:corr} below.

\subsection{BPS spectrum}\label{sec:BPS}

The above example is rather special in that the BPS spectrum of the worldsheet theory (that can be determined in the NS-R formulation) matches exactly that of the dual CFT. This is, to our knowledge, the only example (in the NS-R formulation) where this happens; usually, the NS-R worldsheet theory exhibits some gaps in the BPS spectrum that are not present in the dual CFT, see e.g.\ the analysis in \cite[Sections~4.3 \& 4.4]{Eberhardt:2017pty}. 

It was shown in \cite{Gaberdiel:2018rqv} that the physical worldsheet spectrum reproduces the full (single particle) spectrum of the symmetric orbifold, and it is therefore sufficient to analyse the BPS states in terms of the symmetric orbifold.\footnote{We cannot directly apply the analysis of \cite{Eberhardt:2017pty} since there only states from the discrete representations were considered, whereas now the worldsheet theory only consists of the continuous representations. However, one can also obtain the results below directly from a worldsheet analysis.} First of all, it follows from the analysis of \cite{Gaberdiel:2018rqv}  that the $8$ fermions in the symmetric orbifold transform as $2\cdot ({\bf 2},{\bf 2})$ with respect to the two $\mathfrak{su}(2)$ algebras. By applying the corresponding $-\frac{1}{2}$ modes, we therefore get, in each twisted sector, four large ${\cal N}=4$ BPS states, i.e.\ two large ${\cal N}=4$ multiplets with charges $[j,j,0]$ and $[j+\tfrac{1}{2},j+\tfrac{1}{2},0]$; here the two labels describe the spins with respect to the two $\mathfrak{su}(2)$ algebras, i.e.\ $j^\pm=j$, and we are using the conventions of \cite{Eberhardt:2017pty}.\footnote{As is explained there, each large ${\cal N}=4$ multiplet with charges $[j,j,0]$ contains two BPS states with $\mathfrak{su}(2)$ spins $(j,j)$ and $(j+\frac{1}{2},j+\tfrac{1}{2})$, respectively.} We can therefore concentrate on the `lower' BPS state in each twisted sector. The analysis then depends on whether the twist $w$ is even or odd, and we therefore discuss the two cases in turn. 

For {\bf odd twist} $w$, the ground state of the $w$-cycle twisted sector has conformal dimension
\be
h_0 = (2 B+F) \frac{w^2-1}{48 w} = \frac{w^2-1}{4 w} \ , 
\ee
where we have used the relations from \cite[Appendix~A]{Gaberdiel:2018rqv} and set $B=2$ and $F=8$. The lower BPS state is now obtained from this by applying all $(+\frac{1}{2},+\frac{1}{2})$ charged fermions with mode numbers $-\frac{1}{2} < r <0$. 
There are altogether $2\cdot \frac{w-1}{2}$ such modes --- the overall factor of $2$ comes from the fact that we have two different $(+\frac{1}{2},+\frac{1}{2})$ charged fermions --- and the associated conformal dimension is therefore 
\be
\delta h = 2 \sum_{n=1}^{\frac{w-1}{2}} \bigl( \frac{1}{2} - \frac{n}{w} \bigr) =  \frac{(w-1)^2}{4w} \ .
\ee
Together with the ground state conformal dimension we therefore find
\be
h = h_0 + \delta h = \frac{w^2-1}{4 w}  +  \frac{(w-1)^2}{4w} = \frac{w-1}{2} \ .
\ee
On the other hand, the $\mathfrak{su}(2)\oplus\mathfrak{su}(2)$ spin of this state is $j_+ = j_- = \frac{w-1}{2}$, and the state is therefore indeed BPS. Since $w$ is odd, $j_+ = j_- \in\mathbb{N}_0$, i.e.\ the lower BPS states in each odd cycle twisted sector have integer $\mathfrak{su}(2)$ spin. Together with the global fermion $-\frac{1}{2}$ modes from above, this then gives rise to the BPS multiplets
\be
{\rm BPS}: \qquad \bigoplus_{j=0}^{\infty} \Bigl( [j,j,0] \oplus [j+\tfrac{1}{2},j+\tfrac{1}{2},0] \Bigr) \otimes \overline{ \Bigl( [j,j,0] \oplus [j+\tfrac{1}{2},j+\tfrac{1}{2},0] \Bigr)} \ ,
\ee
where $j$ runs over all non-negative integers. Note that this includes (via the `diagonal' terms) the BPS spectrum expected from supergravity, \be
{\rm sugra BPS}: \qquad \bigoplus_{j=0,\frac{1}{2},1,\ldots} [j,j,0] \otimes \overline{[j,j,0]} \ , 
\ee
but contains also the extra BPS states (the `off-diagonal' terms)
\be
{\rm extra}\ {\rm BPS}: \qquad 
\bigoplus_{j=0,1,2,\ldots} \Bigl[ \bigl( [j+\tfrac{1}{2},j+\tfrac{1}{2},0] \otimes \overline{[j,j,0]}  \bigr)\  \oplus \ \bigl( [j,j,0] \otimes \overline{[j+\tfrac{1}{2},j+\tfrac{1}{2},0]} \bigr) \Bigr]\ . 
\ee
In any case, we note that the only BPS states with $j_\pm=\bar{\jmath}_\pm=\frac{1}{2}$ come from the $w=1$ untwisted sector. In terms of the worldsheet description they arise from the R sector for $j_0=\frac{1}{2}$ and $j_0^\pm =0$, with $w=1$ and $w^\pm=0$, see \cite[Section~4.4.1]{Eberhardt:2017pty}. The fact that there are two such operators is a consequence of the $4$ uncharged fermionic zero modes --- they gives rise to $\frac{1}{2} \cdot 2^2=2$ states after imposing the GSO projection.
\smallskip

For {\bf even twist} $w$, the ground state of the $w$-cycle twisted sector has conformal dimension (for $B=2$ and $F=8$)
\be
h_0 = B \frac{w^2-1}{24 w} + F \frac{w^2+2}{48 w}  = \frac{w^2+1}{4w} \ , 
\ee
see again \cite[Appendix~A]{Gaberdiel:2018rqv}. The negative fermion modes (with mode numbers bigger than $-\frac{1}{2}$) now contribute instead 
\be
\delta h = 2 \sum_{n=1}^{\frac{w-2}{2}} \bigl( \frac{1}{2} - \frac{n}{w} \bigr) =  \frac{w-2}{4} \ , 
\ee
and hence the resulting conformal dimension of the would-be-BPS state is 
\be\label{weven}
h_0 + \delta h = \frac{w^2+1}{4w} + \frac{w-2}{4} = \frac{w-1}{2} + \frac{1}{4w} \ . 
\ee
As before, the $\mathfrak{su}(2)\oplus\mathfrak{su}(2)$ spin of this state is $j_+ = j_- = \frac{w-1}{2}$, and thus the state is not BPS because of the $\frac{1}{4w}$ correction term in (\ref{weven}). Thus we conclude that there are no BPS states in the sectors with $w$ even. Incidentally, this mirrors precisely what happens for the symmetric orbifold of ${\cal S}_0$, see \cite{Eberhardt:2017pty}. 

Because there are no BPS states for even twist, and all the BPS states for odd $w>1$ have spin greater or equal to $+1$, it follows that the only moduli come from the untwisted, i.e.\ the $w=1$ sector. The relevant BPS states with $j_+=j_-=\frac{1}{2}$ are simply the free fermion descendants of the vacuum; the linear combination that corresponds to the fermion descendant of the large ${\cal N}=4$ algebra then leads to the modulus that changes the radius of the compact ${\rm S}^1$. The other modulus is associated to the non-compact boson, resp.\ the $c=1$ Liouville theory. In particular, the latter theory probably possesses a modulus that modifies the reflection coefficient off the Liouville wall.\footnote{We thank Ofer Aharony for a discussion about this point.}

\subsection{Correlation functions}\label{sec:corr}

Since the supersymmetric $\mathfrak{sl}(2,\mathds{R})^{(1)}$ model is at level $k=1$, the correlation functions of the worldsheet theory can be analysed by the same methods as in \cite{Eberhardt:2019ywk}. Relative to the latter theory, the spins now are of the form $j_i = \frac{1}{2} + i s_i$, so the condition that the correlation functions localise, see \cite[eq.~(1.2)]{Eberhardt:2019ywk},
\be
\sum_{i=1}^{n} j_i = \frac{n}{2} \ , 
\ee
is not automatically satisfied, but requires that the  `momenta' $s_i$ sum up to zero, 
\be
\sum_{i=1}^{n} s_i = 0 \ .
\ee
If the $\mathds{R}_0$ theory was just a non-compact boson, this would be the usual momentum conservation condition. 
However, since on the worldsheet, the representation with $j= \frac{1}{2} + is$ is equivalent to that with $j=\frac{1}{2} - is$, the situation is likely to be a bit more complicated, and the theory may rather behave like the $c=1$  Liouville theory, see e.g.\ \cite{Schomerus:2003vv,Runkel:2001ng,McElgin:2007ak,Harlow:2011ny}. In any case, it would be interesting to confirm that a consistent localising worldsheet theory in the NS-R sector can be constructed; this should then also clarify precisely what theory the $c=1$ limit of the linear dilaton theory is. Note that  our $\mathfrak{sl}(2,\mathds{R})_3$ worldsheet theory must be different from the one proposed in \cite{Dei:2021xgh,Dei:2021yom}, but it would not be surprising if for $k=3$ another consistent $\mathfrak{sl}(2,\mathds{R})$ theory exists.\footnote{The analysis of \cite{Dei:2021xgh,Dei:2021yom} only fixes the correlation functions uniquely for generic values of $k$, but for rational $k$, and in particular $k=3$, it is quite possible that a special (and different) solution exists. We thank Lorenz Eberhardt for a discussion about this issue.} After all, given the fact that the ${\rm AdS}_3 \times {\rm S}^3 \times \mathbb{T}^4$ theory is localising at $k=1$, one would expect that this is also visible in terms of the bosonic $\mathfrak{sl}(2,\mathds{R})_3$ worldsheet theory. 

In any case, while the above localisation analysis is maybe a bit inconclusive, we can give a sharper argument for why the correlators must localise in the hybrid description of our theory, see Section~\ref{hybrid} below.

\subsection{Worldsheet supersymmetry}\label{sec:worldsheetsusy}

We close this section by discussing the worldsheet supersymmetry of this background. By construction, the above worldsheet theory has (for any value of the levels) ${\cal N}=1$ superconformal symmetry, see the description in Appendix~\ref{app:conventions}, but one would expect it to have actually ${\cal N}=2$ superconformal symmetry, and this is in fact correct. Indeed, following the classical argument in \cite{Banks:1987cy}, this is expected to follow from the spacetime supersymmetry of the background. Furthermore, the eigenvalues of the ${\cal N}=2$ $\mathfrak{u}(1)$ current are believed to be (half-)integers. However, as had been noted before, for the case of the ${\rm AdS}_3$ backgrounds the situation is somewhat complicated in that the natural ${\cal N}=2$ algebra, say for ${\rm AdS}_3 \times {\rm S}^3 \times \mathbb{T}^4$ \cite{Giveon:1998ns,Israel:2003ry}, does not lead to an (half-)integral $\mathfrak{u}(1)$ spectrum, see e.g.\ \cite{Giveon:1998ns,Ferko:2024uxi}. 

In the following we will argue that for the case of ${\rm AdS}_3 \times {\rm S}^3 \times \mathbb{T}^4$ there is a natural resolution of this small puzzle: as we shall explain, the correct worldsheet algebra is not just the standard ${\cal N}=2$ algebra, but an algebra that contains the ${\cal N}=2$ algebra, as well an additional free boson and fermion. The canonical $\mathfrak{u}(1)$ of this algebra then plays the role of the usual $\mathfrak{u}(1)$ current of \cite{Banks:1987cy}, and its spectrum is indeed (half-)integer. Following a generalisation of the standard construction of \cite{Goddard:1988wv} one can decouple the boson and fermion and then land on the familiar ${\cal N}=2$ algebra of \cite{Giveon:1998ns,Israel:2003ry}. However, in the process the $\mathfrak{u}(1)$ gets modified, and the resulting $\mathfrak{u}(1)$ generator does not have an half-integral spectrum any longer. We will then also present the corresponding analysis for the case of ${\rm AdS}_3 \times {\rm S}^3 \times {\rm S}^3 \times {\rm S}^1$ where the situation is analogous. 

\subsubsection[A brief review of the analysis for \texorpdfstring{${\rm AdS}_3 \times {\rm S}^3 \times \mathbb{T}^4$}{AdS3xS3xT4}]{\boldmath A brief review of the analysis for \texorpdfstring{${\rm AdS}_3 \times {\rm S}^3 \times \mathbb{T}^4$}{AdS3xS3xT4}}

Starting from the ${\cal N}=1$ generators of the full background, we can obtain the ${\cal N}=2$ generators by identifying the $\mathfrak{u}(1)$ current $J$ of the putative ${\cal N}=2$ theory. Following the construction of \cite{Banks:1987cy} the exponential of $J$ will give rise to the spacetime supercharge, and for the case of ${\rm AdS}_3 \times{\rm S}^3 \times \mathbb{T}^4$, the relevant current is  \cite{Giveon:1998ns,Ferko:2024uxi} 
\begin{equation}
J= \psi^+\psi^- +\chi^+\chi^-+2\psi^3\chi^3-2i\lambda^1\lambda^2+2i\lambda^3\lambda^4\ , 
\label{eq:JT4}
\end{equation}
where we are using the conventions of Appendix~\ref{app:conventions}, and the fermions of the torus $\mathbb{T}^4$ are denoted by $\lambda^i$ with OPEs 
\be
\lambda^i(z) \lambda^j(w) \sim \frac{\delta^{ij}}{2(z-w)} \ . 
\ee
The associated ${\cal N}=2$ generators are then (here $k$ is the level of both the superconformal $\mathfrak{sl}(2,\mathds{R})$ and $\mathfrak{su}(2)$ algebra)
\begin{equation}
\begin{aligned}
G^+=&\sqrt{\frac{2}{k}}\Bigl[ \mathcal{J}^-\psi^+-\mathcal{J}^3(\chi^3+\psi^3)-\psi^+\psi^-(\chi^3+\psi^3)+\mathcal{K}^-\chi^++\mathcal{K}^3(\chi^3+\psi^3) \nonumber \\
& \qquad +\chi^+\chi^-(\chi^3+\psi^3)\Bigr] + G^+_{\mathbb{T}^4}\\
G^-=&\sqrt{\frac{2}{k}}\Bigl[ \mathcal{J}^+\psi^-+\mathcal{J}^3(\chi^3-\psi^3)+\psi^+\psi^-(\chi^3-\psi^3)+\mathcal{K}^+\chi^-+\mathcal{K}^3(\chi^3-\psi^3)\nonumber \\ 
& \qquad +\chi^+\chi^-(\chi^3-\psi^3)\Bigr] + G^-_{\mathbb{T}^4}\ ,
\end{aligned}
\end{equation}
where $G^\pm_{\mathbb{T}^4}$ are the usual ${\cal N}=2$ generators associated to $\mathbb{T}^4$. The complication relative to the usual case now arises in that the supercharges $G^\pm$ and $J$ do not define a standard ${\cal N}=2$ algebra with $c=15$, but involve an additional free boson and fermion. More concretely, we find
\begin{equation}
\begin{array}{rclrcl}
{}[J_m,J_n] & = & 5 \, m \, \delta_{m,-n} &  &   &    \\ 
{}[J_m,G^+_r] & = & G^+_{m+r} \ , \qquad & {}[J_m,G^-_r] & = & - G^-_{m+r} +  m \, \Psi_{m+r} \\
{}[J_m,U_n] & = & 0 \ , \qquad & [J_m,\Psi_r] & = & - \Psi_{m+r} \\
\{ G^+_r,\Psi_{s}\} & = & -2 U_{r+s} \ , \qquad
& [U_m,G^+_r] & = & 0 \ , \\
\{ G^-_r,\Psi_{s}\} & = &0 \ , \qquad
& [U_m, G^-_r] & = & -m\, \Psi_{m+r} \ ,  \label{2.12}\\
{}[L_m,J_n] & = & - n J_{m+n} \ , \qquad & [L_m,U_n] & = & - n U_{m+n} \ , \\
{}[L_m,G^\pm_r] & = & (\frac{m}{2}-r) G^\pm_{m+r} \ , \qquad & [L_m, \Psi_r] & = & (-\frac{m}{2}-r) \Psi_{m+r} \ ,
\end{array}
\end{equation}
as well as 
\be  \label{2.13}
\begin{aligned}
\{G^+_r, G^-_s\}  & = 2L_{r+s} + (r-s) (J_{r+s} + U_{r+s} )  + 5 \, (r^2-\tfrac{1}{4}) \delta_{r,-s} \ , \\
{}[L_m,L_n] & = (m-n) L_{m+n} + \tfrac{15}{12} m (m^2-1) \, \delta_{m,-n} \ . 
\end{aligned}
\ee
Here $U$ and $\Psi$ are the additional boson and fermion, and they are given explicitly by 
\be\label{2.15}
U=\tfrac{2}{k} \bigl( \mathcal{J}^3-\mathcal{K}^3+\psi^+\psi^--\chi^+\chi^- \bigr)\ ,\qquad \Psi= \tfrac{2\sqrt{2}}{\sqrt{k}} \bigl( \chi^3-\psi^3 \bigr)\ .
\ee
Both of them are `null' in the sense that\footnote{This does not mean, however, that they are `null' fields in the usual sense since, for example, $\{\chi^3_r,\Psi_s\} = \sqrt{\frac{2}{k}}\,\delta_{r,-s}$.}
\be
{}[U_m,U_n]=0 \ , \qquad \{ \Psi_r,\Psi_s\} = 0 \ ,
\ee
and, as a consequence, one cannot quite follow the prescription of \cite{Goddard:1988wv}. However, the above algebra contains a standard ${\cal N}=2$ subalgebra generated by $L_m$, $G^\pm_r$ and $V_n$ (instead of $J_n$), where 
\be\label{Vdef}
V = J + U \ ,
\ee
and in fact this just reproduces the original construction of \cite{Israel:2003ry}. Because of the $U$ correction term, the spectrum of $V$ is then not in general (half-)integer. Note that $U$ can be thought of as a curvature correction since it is proportional to $\frac{1}{k}$. One can also check that the corresponding supercharges are indeed BRST invariant, and this is briefly explained in Appendix~\ref{A.2.1}.

\subsubsection[The construction for \texorpdfstring{${\rm AdS}_3\times {\rm S}^3 \times {\rm S}^3 \times {\rm S}^1$}{AdS3xS3xS3xS1}]{\boldmath The construction for \texorpdfstring{${\rm AdS}_3\times {\rm S}^3 \times {\rm S}^3 \times {\rm S}^1$}{AdS3xS3xS3xS1}}

The construction for ${\rm AdS}_3\times {\rm S}^3 \times {\rm S}^3 \times {\rm S}^1$ proceeds similarly. The analogue of the current $J$ above is now 
\begin{align}
J &=\psi^+\psi^-+\chi^{(+)+}\chi^{(+)-}+\chi^{(-)+}\chi^{(-)-}  \label{eq:JS3xS1}  \\ 
& +2\psi^3\left(\sqrt{\tfrac{k}{k_+}}\chi^{(+)3}+\sqrt{\tfrac{k}{k_-}}\chi^{(-)3}\right)
+2i\lambda\left(\sqrt{\tfrac{k}{k_-}}\chi^{(+)3}-\sqrt{\tfrac{k}{k_+}}\chi^{(-)3}\right)\ , \nonumber
\end{align}
and the associated supercurrents are 
\begin{align}
G^+=&\sqrt{\frac{2}{k}}\Bigl[( \mathcal{J}^-\psi^+-(\mathcal{J}^3+\psi^+\psi^-)\left(\psi^3+\left(\sqrt{\tfrac{k}{k_+}}\chi^{(+)3}+\sqrt{\tfrac{k}{k_-}}\chi^{(-)3}\right)\right) \Bigr] \nonumber\\
&+\sqrt{\frac{2}{k_+}}\Bigl[ \mathcal{K}^{(+)-}\chi^{(+)+}+(\mathcal{K}^{(+)3}+\chi^{(+)+}\chi^{(+)-})\left(\sqrt{\tfrac{k}{k_+}}\psi^3+i\sqrt{\tfrac{k}{k_-}}\lambda+\chi^{(+)3}\right) \Bigr] \nonumber \\
&+\sqrt{\frac{2}{k_-}}\Bigl[ \mathcal{K}^{(-)-}\chi^{(-)+}+(\mathcal{K}^{(-)3}+\chi^{(-)+}\chi^{(-)-})\left(\sqrt{\tfrac{k}{k_-}}\psi^3-i\sqrt{\tfrac{k}{k_+}}\lambda+\chi^{(-)3}\right) \Bigr] \nonumber \\
&+\partial\phi\Bigl(\lambda-i\left(\sqrt{\tfrac{k}{k_-}}\chi^{(+)3}-\sqrt{\tfrac{k}{k_+}}\chi^{(-)3}\right) \ ,  \Bigr)\\
G^-=&\sqrt{\frac{2}{k}}\Bigl[ \mathcal{J}^+\psi^--(\mathcal{J}^3+\psi^+\psi^-)\left(\psi^3-\left(\sqrt{\tfrac{k}{k_+}}\chi^{(+)3}+\sqrt{\tfrac{k}{k_-}}\chi^{(-)3}\right)\right) \Bigr] \nonumber \\
&+\sqrt{\frac{2}{k_+}} \Bigl[ \mathcal{K}^{(+)+}\chi^{(+)-}+(\mathcal{K}^{(+)3}+\chi^{(+)+}\chi^{(+)-})\left(-\sqrt{\tfrac{k}{k_+}}\psi^3-i\sqrt{\tfrac{k}{k_-}}\lambda+\chi^{(+)3}\right) \Bigr] \nonumber \\
&+\sqrt{\frac{2}{k_-}}\Bigl[ \mathcal{K}^{(-)+}\chi^{(-)-}+(\mathcal{K}^{(-)3}+\chi^{(-)+}\chi^{(-)-})\left(-\sqrt{\tfrac{k}{k_-}}\psi^3+i\sqrt{\tfrac{k}{k_+}}\lambda+\chi^{(-)3}\right) \Bigr] \nonumber \\
&+\partial\phi\Bigl(\lambda+i\left(\sqrt{\tfrac{k}{k_-}}\chi^{(+)3}-\sqrt{\tfrac{k}{k_+}}\chi^{(-)3}\right)  \Bigr)\ . 
\end{align}
They then generate an extended ${\cal N}=2$ algebra as above, see eqs.~(\ref{2.12}) and (\ref{2.13}), where now\footnote{As a sanity check one finds that the $k_-\to\infty$ limit of (\ref{2.21}) and (\ref{2.22}) reproduces (\ref{2.15}) above.}
\begin{align}
\Psi=&\, 2\sqrt{2}\left(\frac{\chi^{(+)3}}{\sqrt{k_+}}+\frac{\chi^{(-)3}}{\sqrt{k_-}}-\frac{\psi^{3}}{\sqrt{k}}\right) \label{2.21}\\
U=&\, 2\left(\frac{\mathcal{J}^3+\psi^+\psi^-}{k}-\frac{\mathcal{K}^{(+)3}+\chi^{(+)+}\chi^{(+)-}}{k_+}-\frac{\mathcal{K}^{(-)3}+\chi^{(-)+}\chi^{(-)-}}{k_-}\right) \ . \label{2.22}
\end{align}
As before, this reduces to a conventional ${\cal N}=2$ algebra with $c=15$ provided we replace $J$ by $V= J+U$, see eq.~(\ref{Vdef}). The spectrum of $J$ is (half-)integer for any choice of the levels, but that of $U$ generically is not. We have also checked that the corresponding supercharges are indeed BRST invariant, see Appendix~\ref{A.2.2}.

\section{The hybrid approach}\label{hybrid}

In this section we discuss the hybrid description of the theory with $k_\pm=2$ and $k=1$. In particular, we shall find that it possesses a free field realisation from which the desired localisation properties of Section~\ref{sec:corr} can be deduced following essentially the same arguments as in \cite{Dei:2020zui}.

To describe the hybrid version of our model we follow the analysis of \cite{Eberhardt:2019niq}. As is explained in Section~3.2 of that paper, in the hybrid approach the worldsheet fields consist of, see  \cite[eq.~(3.6)]{Eberhardt:2019niq}\footnote{We will use the convention that the `level' $k$ of $\mathfrak{d}(2,1;\alpha)$ is the level of the $\mathfrak{sl}(2,\mathds{R})$ factor.}
\be
\mathfrak{d}(2,1;\alpha)_k \oplus \mathfrak{u}(1) \oplus \hbox{2 pairs of top.\ twisted fermions} \oplus bc \ \hbox{and} \ \rho \ \hbox{ghosts} \ . 
\ee
In \cite{Eberhardt:2019niq} the representations for which one of the two $\mathfrak{su}(2)$ levels is one, say $k_+ =1$, were analysed, and it was suggested that this is the moral analogue of the situation analysed in \cite{Eberhardt:2019ywk,Eberhardt:2018ouy}. In view of our discussion above, we now want to concentrate instead on the situation described by (\ref{tensionless}). In that case, there will be generically long multiplets of $\mathfrak{d}(2,1;\alpha)_1$, as we shall now explain. 

\subsection[Analysis of the \texorpdfstring{$\mathfrak{d}(2,1|\alpha)_1$}{d(2,1|a)1} representations]{\boldmath Analysis of the \texorpdfstring{$\mathfrak{d}(2,1|\alpha)_1$}{d(2,1|a)1} representations}

Following the conventions of \cite{Eberhardt:2019niq} the generic long representation has the form\be 
\begin{tabular}{ccccccccc}
& $(\mathcal{C}^{j-1}_{\alpha},\mathbf{m}^+,\mathbf{m}^-)$ & \\
& $(\mathcal{C}^{j-\frac{1}{2}}_{\alpha+\frac{1}{2}},\mathbf{m^+\pm 1},\mathbf{m^-\pm 1})$ & \\
$(\mathcal{C}^{j}_{\alpha},\mathbf{m^+\pm 2},\mathbf{m^-})$ & $2 \cdot (\mathcal{C}^{j}_{\alpha},\mathbf{m^+},\mathbf{m^-})$ & $(\mathcal{C}^{j}_{\alpha},\mathbf{m^+},\mathbf{m^-\pm 2})$ \\
& $(\mathcal{C}^{j+\frac{1}{2}}_{\alpha+\frac{1}{2}},\mathbf{m^+\pm 1},\mathbf{m^-\pm 1})$ & \\
& \phantom{\ ,}$(\mathcal{C}^{j+1}_{\alpha},\mathbf{m^+},\mathbf{m^-}) \ ,$   &
\end{tabular}\label{eq:long representation} 
\ee
\pagebreak

\noindent where $\mathcal{C}^{j}_{\alpha}$ denotes the $\mathfrak{sl}(2,\mathds{R})_1$ representation with spin $j$ and $J^3_0$ eigenvalues $\alpha +\mathbb{Z}$.\footnote{Note that this `$\alpha$' parameter is not to be confused with the $\alpha$ parameter appearing in $\mathfrak{d}(2,1|\alpha)_1$; for us the latter $\alpha$ parameter always equals $\alpha=1$, see eq.~(\ref{alpha}).}

At $k_\pm =2$ only  $\mathfrak{su}(2)^{(\pm)}$ representations with dimension less than or equal to ${\bf 3}$ are allowed, and hence we need to choose $\mathbf{m}^\pm = {\bf 1}$ in the first line, and impose the obvious truncations that are a consequence of the fact that only the $({\bf 2},{\bf 2})$ representation appears in the second line; this then leads to the long representation 
\begin{equation}
\begin{gathered}
(\mathcal{C}^{j-1}_{\alpha},\textbf{1},\textbf{1})\\
(\mathcal{C}^{j-\frac{1}{2}}_{\alpha-\frac{1}{2}},\textbf{2},\textbf{2})\\
(\mathcal{C}^{j}_{\alpha},\textbf{3},\textbf{1})\qquad (\mathcal{C}^{j}_{\alpha},\textbf{1},\textbf{3}) \\
(\mathcal{C}^{j+\frac{1}{2}}_{\alpha-\frac{1}{2}},\textbf{2},\textbf{2})\\
(\mathcal{C}^{j+1}_{\alpha},\textbf{1},\textbf{1})
\label{eq: long repn in d(2,1|alpha)}
\end{gathered} \ .
\end{equation}
Since there is no `shortening', this representation exists for all values of the $\mathfrak{sl}(2,\mathds{R})$ spin, and we will postulate that $j$ will parametrise an arbitrary continuous representation. As in the NS-R description above, this will lead to a modular invariant partition function on the worldsheet; this will be shown below, once we have introduced a convenient free field realisation for this background. The value of the $\mathfrak{d}(2,1|\alpha)$ Casimir 
\begin{align}
\mathcal{C}^{\mathfrak{d}(2,1;\alpha)}&=\mathcal{C}_\text{bos}^{\mathfrak{d}(2,1;\alpha)}+\mathcal{C}_\text{ferm}^{\mathfrak{d}(2,1;\alpha)}\ , \\
\mathcal{C}_\text{bos}^{\mathfrak{d}(2,1;\alpha)}&=\mathcal{C}^{\mathfrak{sl}(2,\mathds{R})}+\gamma\mathcal{C}^{\mathfrak{su}(2)_+}+(1-\gamma)\mathcal{C}^{\mathfrak{su}(2)_-}\ , \\
\mathcal{C}_\text{ferm}^{\mathfrak{d}(2,1;\alpha)}&=-\frac{1}{2} \epsilon^{\alpha\mu} \epsilon^{\beta\nu}\epsilon^{\gamma \rho} S_0^{\alpha\beta\gamma} S_0^{\mu\nu\rho}
\end{align}
on this representation turns out to equal $\frac{1}{4} + s^2$, where we have written $j=\frac{1}{2} + is$. In order to see this, we evaluate it on the state $\psi$ in $(\mathcal{C}^{j}_{\alpha},\textbf{1},\textbf{3})$ that has $K^{(-)3}_0$ eigenvalue $+1$, and hence is annihilated by 
\be
S_0^{\alpha\beta+} \, \psi =0 \ . 
\ee
Then the fermionic part of the quadratic Casimir becomes 
\begin{equation}
\begin{aligned}
-\frac{1}{2}\epsilon^{\alpha\mu}\epsilon^{\beta\nu}\epsilon^{\gamma\rho}S^{\alpha\beta\gamma}_0S^{\mu\nu\rho}_0\, \psi=&-\frac{1}{2}\epsilon^{\alpha\mu}\epsilon^{\beta\nu}S^{\alpha\beta+}_0S^{\mu\nu-}_0\, \psi=-\frac{1}{2}\epsilon^{\alpha\mu}\epsilon^{\beta\nu}\left\{ S^{\alpha\beta+}_0,S^{\mu\nu-}_0 \right\}\, \psi\\
=&-K^{(-)3}_0\, \psi =-\psi\ ,
\end{aligned}
\end{equation}
and the full $\mathfrak{d}(2,1|\alpha)$ Casimir is therefore 
\begin{equation}
\mathcal{C}^{\mathfrak{d}(2,1;\alpha)} = -\left(\frac{1}{2}+is\right)\left( \frac{1}{2}-1+is \right)+\frac{1}{2}\times0(0+1)+\frac{1}{2}\times1(1+1) - 1 =\frac{1}{4}+s^2\ .
\end{equation}

\subsection{Free field realisation}\label{sec:free}

The localisation analysis in \cite{Dei:2020zui} relied on a free field realisation of $\mathfrak{psu}(1,1|2)_1$ in terms of two  pairs of symplectic bosons (and four fermions),\footnote{The localisation analysis essentially only depends on the symplectic bosons.} and one may therefore expect that also in our case such a free field realisation should exist. This turns out to be true, as we shall now explain.

For the case of $\mathfrak{d}(2,1;\alpha)_1$ we again need two pairs of symplectic bosons $\xi^\alpha$  and $\bar\xi^\alpha$ where $\alpha=\pm$, whose OPEs we take to be
\begin{equation}
\bar\xi^\alpha(z)\, \xi^\beta(w)\sim\frac{\epsilon^{\alpha\beta}}{z-w}\ ,
\end{equation}
where $\epsilon^{+-}=-\epsilon^{-+}=1$. In addition we now have $8$ real fermions $\psi_i^{\alpha\beta}$, where $\alpha,\beta=\pm$ and $i=1,2$, with OPEs
\begin{equation}
\begin{aligned}
\psi^{\alpha\beta}_i(z)\, \psi^{\mu\nu}_j(w)\sim&-\frac{\delta_{ij}\epsilon^{\alpha\mu}\epsilon^{\beta\nu}}{z-w}\ .
\end{aligned}
\end{equation}
Then the $\mathfrak{sl}(2,\mathds{R})$ currents are given by
\begin{equation}
\begin{gathered}
J^3=-\frac{1}{2}(\xi^+\bar\xi^-+\xi^-\bar\xi^+)\ ,\quad J^\pm=\xi^\pm\bar\xi^\pm\ ,
\end{gathered}
\end{equation}
while the two $\mathfrak{su}(2)_2$ currents are (repeated indices are summed over)
\begin{equation}
\begin{gathered}
K^{(\pm)3}=-\frac{1}{2}(\psi^{++}_i\psi^{--}_i\pm\psi^{-+}_i\psi^{+-}_i)\ ,\quad 
K^{(+)\pm}=\pm\psi^{\pm+}_i\psi^{\pm-}_i,\quad K^{(-)\pm}=\pm\psi^{+\pm}_i\psi^{-\pm}_i\ .
\end{gathered}
\end{equation}
Finally, the supercurrents take the form
\begin{equation}
S^{\alpha\beta\gamma}=\frac{1}{2}\left(\xi^\alpha(\psi_1+i\psi_2)^{\beta\gamma}-\bar\xi^\alpha(\psi_1-i\psi_2)^{\beta\gamma}\right)\ .
\label{eq:supercharges}
\end{equation}
It is not difficult to check that the corresponding modes satisfy then indeed the relations of eqs.~(\ref{eq:d21alpha}) with 
\be
k=1 \ , \qquad k_\pm = 2 \ , \qquad \gamma = \tfrac{1}{2}   \ . 
\ee
The total central charge of the free fields is 
\be
c_{\rm free} = 4 \times (-\tfrac{1}{2}) + 8 \times (+\tfrac{1}{2}) = 2 \ , 
\ee
whereas the central charge of $\mathfrak{d}(2,1|\alpha)$ is $c \bigl(\mathfrak{d}(2,1|\alpha)\bigr)=1$. Thus we should expect that the commutant of $\mathfrak{d}(2,1|\alpha)$ contains a $\mathfrak{u}(1)$ field $Z$, and $Z$ is indeed explicitly given by 
\be
Z = -2U + V \ , 
\ee
where
\be
U=-\frac{1}{2}\xi^+\bar\xi^-+\frac{1}{2}\xi^-\bar\xi^+\ , \qquad V=i\epsilon^{\alpha\beta}\epsilon^{\mu\nu}\psi_1^{\alpha\mu}\psi_2^{\beta\nu} \ . 
\ee
Here $V$ has the property that the fermions $(\psi_1\pm i\psi_2)^{\beta\gamma}$ have charge $\pm 1$ with respect to $V$. Unlike the situation for $\mathfrak{psu}(1,1|2)_1$, $Z$ is however not `null', but rather has the OPE 
\begin{equation}
Z(z)\, Z(w)\sim\frac{2}{(z-w)^2}\ .
\end{equation}
Here the coefficient in the numerator arises as $-2+4$, where $-2$ comes from the symplectic bosons, while the $4$ is the contribution from the fermions. Since $Z$ is not null, we should expect that $\mathfrak{d}(2,1|\alpha)_1$ is simply the coset of the free field theory by $Z$.\footnote{Our arguments so far only show that $\mathfrak{d}(2,1|\alpha)_1$ is contained in this coset, but it is conceivable that the coset algebra is strictly bigger (although we have not found any evidence for this).} In fact, this precisely fits together with the coset formula for the stress energy tensor since the free field stress energy tensor  
\be
T_{\rm free} = -\frac{\epsilon^{\alpha\beta}}{2}(\partial\bar\xi^\alpha\xi^\beta+\partial\xi^\alpha\bar\xi^\beta)-\frac{1}{2}\epsilon^{\beta\nu}\epsilon^{\gamma\rho}\partial\psi^{\beta\gamma}_i\psi^{\nu\rho}_i\ , 
\ee
equals exactly the sum 
\be\label{3.21}
T_{\rm free} = T^{\mathfrak{d}(2,1|\alpha)_1} + \frac{Z^2}{4} \ , 
\ee
where the $\mathfrak{d}(2,1|\alpha)_1$ stress tensor is given by  
\begin{align}
T^{\mathfrak{d}(2,1|\alpha)_1}   = &  -(J^3)^2+\frac{1}{2}(J^+J^-+J^-J^+) -\frac{1}{2}\epsilon^{\alpha\mu}\epsilon^{\beta\nu}\epsilon^{\gamma\rho}S^{\alpha\beta\gamma}S^{\mu\nu\rho} \nonumber \\
& +  \frac{1}{4}\Bigl( 2(K^{(+)3})^2 + (K^{(+)+}K^{(+)-}+K^{(+)-}K^{(+)+}) \Bigr) \nonumber \\ 
& + \frac{1}{4}\Bigl( 2(K^{(-)3})^2 + (K^{(-)+}K^{(-)-}+K^{(-)-}K^{(-)+}) \Bigr)  \ . 
\end{align}

\subsection{The long representation in the free field realisation}

It is not difficult to describe the above long representation, see eq.~(\ref{eq: long repn in d(2,1|alpha)}), in terms of the  free field construction of the previous section. Following \cite{Dei:2020zui}, we label the bosonic highest weight states as $|m_1,m_2\rangle$, where 
\be\label{xizero}
\begin{array}{rclrcl}
\bar\xi^+_0 |m_1,m_2\rangle & = &  \, |m_1,m_2+\tfrac{1}{2}\rangle \ , \qquad 
&
\xi^+_0 |m_1,m_2\rangle & = & 2 \, m_1 \, |m_1+\tfrac{1}{2},m_2\rangle \  , \\[3pt]
\bar\xi^-_0 |m_1,m_2\rangle & = & - \, |m_1-\tfrac{1}{2},m_2\rangle \ , \qquad 
& {\xi}^-_0 |m_1,m_2\rangle & = & - 2 \, m_2 \, |m_1,m_2-\tfrac{1}{2}\rangle \ .
\end{array}
\ee
The $J^3_0$ and $U_0$ eigenvalues are then 
\begin{eqnarray} \label{eigenvs}
J^3_0 \, |m_1,m_2\rangle & =& (m_1+m_2) \, |m_1,m_2\rangle \label{J30} \\
U_0 \, |m_1,m_2\rangle & =& (m_1-m_2 - \tfrac{1}{2}) \, |m_1,m_2\rangle  \ , \label{U0}
\end{eqnarray}
and the spin of the $\mathfrak{sl}(2,\mathds{R})$ representation is 
\be
j = m_1 - m_2 \ , 
\ee
see \cite[eq.~(2.16)]{Dei:2020zui}. For the action of the fermionic zero modes we take the annihilation operators to be 
\begin{equation}\label{42}
(\psi^{\alpha+}_i)_0 \, |m_1,m_2\rangle=0\ ,
\end{equation}
which implies that the state $|m_1,m_2\rangle$ has $K^{(\pm)3}_0$ eigenvalues 
\be
K^{(+)3}_0\, |m_1,m_2\rangle=0 \ , \qquad K^{(-)3}_0\, |m_1,m_2\rangle= + 1 \, |m_1,m_2\rangle \ , 
\ee
i.e.\ it transforms as the $m=1$ component of $({\bf 1},{\bf 3})$. Furthermore, the state $|m_1,m_2\rangle$ has $V_0$ eigenvalue zero, $V_0 \, |m_1,m_2\rangle = 0$. Applying the fermionic creation operators 
$(\psi_1\pm i\psi_2)^{\alpha -}$, this then generates the states in the $\mathfrak{su}(2)_\pm$ representations
\be
({\bf 3},{\bf 1}) \oplus ({\bf 1},{\bf 3}) \oplus 2\cdot ({\bf 2},{\bf 2}) \oplus 2\cdot ({\bf 1},{\bf 1}) \ . 
\ee
Given that these creation operators have $V$ eigenvalues $\pm 1$, the generators of $\mathfrak{d}(2,1|\alpha)$ (that commute with $Z$), must carry the corresponding opposite $U$ eigenvalues. This is then responsible for shifting the spin $j$ along the diamond, see eq.~(\ref{eq: long repn in d(2,1|alpha)}). 

\subsection{Worldsheet spectrum and localisation}

With these preparations we can now describe the worldsheet spectrum of the hybrid theory. The analogue of 
the NS-R partition function (\ref{wsspec}) is the partition function 
\be\label{hybrid_p}
{\cal H} =  \Bigl[ \sum_{w\in\mathbb{Z}} \int_0^1d \alpha \int_{\mathds{R}} ds\, \sigma^w \big( {\cal R}(s, \alpha) \bigr) \otimes \sigma^w \bigl( \overline{{\cal R}(s, \alpha)} \bigr) \Bigr] \otimes {\cal H}^{S^1}  \ ,
\ee
where ${\cal R}( s, \alpha)$ denotes the coset of the above free field realisation by the $\mathfrak{u}(1)$ subalgebra generated by $Z$ --- as explained in eq.~(\ref{3.21}) only this coset has the correct stress tensor. Here $j=\frac{1}{2} + is$ labels the spin of the state in the $({\bf 1},{\bf 3})$ representation, while $\alpha$ describes the $J^3_0$ eigenvalue mod integers; since on the state $|m_1,m_2\rangle$ we have the eigenvalues $U_0=i s$ and $V_0=0$, this representation consists then of all free field states that are annihilated by $Z_n$ with $n\geq 0$ and have $Z_0$ eigenvalue equal to $Z_0 = - 2 i s$. Finally, spectral flow acts on the free fields as 
\be\label{specflow}
\sigma^w\bigl((\psi_i^{\alpha\beta})_r\bigr) = (\psi_i^{\alpha\beta})_{r+\frac{\alpha w}{2}}\ , \qquad 
\sigma^w(\xi^{\alpha}_r)= \xi^{\alpha}_{r-\frac{\alpha w}{2}}\ , \qquad 
\sigma^w(\bar\xi^{\alpha}_r)= \bar\xi^{\alpha}_{r-\frac{\alpha w}{2}}\ .
\ee
This then induces the familiar action on the $\mathfrak{d}(2,1|\alpha)$ generators, see \cite{Eberhardt:2019niq}, 
\begin{equation}
\begin{array}{rclrcl}
\sigma^w(K^{(+)3}_n) & = & K^{(+)3}_n+w\delta_{n,0}\ ,\qquad & \sigma^w(K^{(+)\pm}_n) & = & K^{(+)\pm}_{n\pm w}\ , \\
\sigma^w(J^{3}_n) & =& J^{3}_n+\frac{w}{2}\delta_{n,0}\ ,\qquad & \sigma^w(J^{\pm}_n) & = & J^{\pm}_{n\mp w}\ , \\
\sigma^w(K^{(-)a}_n) & = & K^{(-)a}_n \ , \qquad & \sigma^w(L_0) & = & L_0+w(K^{(+)3}_0-J^3_0)+\frac{w^2}{4}\ ,
\end{array}
\end{equation}
while $Z_n$ is invariant. This worldsheet spectrum is then modular invariant, see Appendix~\ref{app:hybrid} for more details. 

In terms of this free field realisation we can then repeat directly the localisation analysis of \cite{Dei:2020zui}, and since this only involved the symplectic bosons, it works again as in \cite{Dei:2020zui}. This therefore gives strong support to the claim that the worldsheet correlators localise as expected for a symmetric orbifold theory (without any deformation).

\section{Conclusions}\label{sec:conc}

In this paper we have given strong support to the assertion that string theory on ${\rm AdS}_3\times {\rm S}^3 \times {\rm S}^3 \times {\rm S}^1$ with one unit of NS-NS flux through the ${\rm AdS}_3$ (and two units of NS-NS flux through each of the two $3$-spheres) is exactly dual to the symmetric orbifold of eight free fermions and two bosons, see eq.~(\ref{1.7}) above. (This is a small modification of what was originally proposed in \cite{Gaberdiel:2018rqv}, see also \cite{Giribet:2018ada}.) In particular, we have shown that the spectrum matches exactly. Furthermore, in the hybrid version of the theory, the correlators exhibit the expected localisation properties using the free field realisation of Section~\ref{sec:free} and the arguments of \cite{Dei:2020zui}. Finally, we noted that there are no exactly marginal large ${\cal N}=4$ moduli in any of the twisted sectors of the symmetric orbifold, and hence no candidate for a deforming field as in \cite{Eberhardt:2021vsx}. (The only exactly marginal ${\cal N}=4$ moduli come from the untwisted sector and hence deform the seed theory.) 

While the localisation is essentially manifest in the hybrid description, it would be very interesting to confirm it directly in the NS-R formulation; this amounts to finding a consistent (crossing-symmetric) $\mathfrak{sl}(2,\mathds{R})_3$ theory that localises to the covering maps. (The idea that such a theory should exist also seems natural from the perspective of the ${\rm AdS}_3 \times {\rm S}^3 \times \mathbb{T}^4$ theory at $k=1$.) It would also be interesting to understand whether the free field realisation that we found in the hybrid formalism, see Section~\ref{sec:free}, has an analogue along the lines of \cite{Dei:2023ivl,Beem:2023dub}, i.e.\ without the need to perform a coset by a $\mathfrak{u}(1)$.
\smallskip

Our example has a number of nice properties relative to the more familiar case of the ${\rm AdS}_3 \times {\rm S}^3 \times \mathbb{T}^4$ theory at $k=1$. First of all, one can describe the theory in the NS-R formalism, without the need to introduce the technically more complicated hybrid description. The other notable difference is that now not just short representations appear in the worldsheet spectrum; this seems to suggest that this `topological' aspect of the duality may not be crucial in general. In any case, it would be interesting to understand this theory (and its deformations) in more detail; for example, it would be instructive and illuminating to extend the analysis of \cite{Aharony:2024fid} (who studied the moduli space for the case of ${\rm AdS}_3 \times {\rm S}^3 \times \mathbb{T}^4$) to this class of backgrounds, see also \cite{Gukov:2004ym}.

.

\acknowledgments
We thank Ofer Aharony, Nathan Benjamin, Minjae Cho, Clay Cordova, Andrea Dei, Lorenz Eberhardt, Rajesh Gopakumar, Egor Im, Bob Knigh\-ton, David Kutasov, Emil Martinec, Jacob McNamara, Kiarash Naderi, Beat Nairz, Hirosi Ooguri, Cheng Peng, Massimo Porrati, Savdeep Sethi, Nick Warner, and Zhe-fei Yu for useful conversations, and in particular Lorenz Eberhardt and Bob Knighton for comments on a first draft. VS would like to thank the High Energy Theory group at the University of Southern California, the Walter Burke Institute for Theoretical Physics at Caltech, the Kadanoff Center for Theoretical Physics at the University of Chicago, and the Princeton Center for Theoretical Science for hospitality during the final stages of this work.
The work of VS is supported through a personal grant of MRG from the Swiss National Science Foundation, and the work of the group at ETH is also supported in part by the Simons Foundation grant 994306  (Simons Collaboration on Confinement and QCD Strings), as well as the NCCR SwissMAP that is also funded by the Swiss National Science Foundation.

\appendix

\section{NS-R conventions}\label{app:conventions}

We shall be using the same conventions for $\mathfrak{sl}(2,\mathbb{R})^{(1)}$ and $\mathfrak{su}(2)^{(1)}$ as in \cite{Ferreira:2017pgt}, except that we have rescaled the fermions. For the supersymmetric $\mathfrak{sl}(2,\mathbb{R})^{(1)}$ algebra we consider 
\begin{alignat}{5}
\bigl[J^{+}_{m},J^{-}_{n}\bigr] 
={}&
 -2J^{3}_{m+n} + km\delta_{m,-n}&
 & \quad &
 \bigl[J^{3}_{m},J^{\pm}_{n}\bigr] 
={}&
 \pm J^{\pm}_{m+n}&
  & \quad &
 \bigl[J^{3}_{m},J^{3}_{n}\bigr] 
={}&
 -\frac{k}{2}m\delta_{m,-n}
 \nonumber\\
 \bigl[J^{\pm}_{m},\psi^{3}_{r}\bigr]
={}&
 \mp\psi^{\pm}_{m+r} &
 & \quad &
 \bigl[J^{3}_{m},\psi^{\pm}_{r}\bigr]
={}&
 \pm\psi^{\pm}_{m+r}&
  &\quad &
 \bigl[J^{\pm}_{m},\psi^{\mp}_{r}\bigr]
 ={}&
 \mp 2\psi^{3}_{m+r}
\nonumber  \\
 \bigl\{\psi^{+}_{r},\psi^{-}_{s}\bigr\}
={}&
 \delta_{r,-s}&
 & \quad &
 \bigl\{\psi^{3}_{r},\psi^{3}_{s}\bigr\}
={}&
 -\frac{1}{2}\delta_{r,-s}\ .  &
\end{alignat}
We can decouple the bosons from the fermions by defining the generators 
\begin{align}
\cJ^{+} 
={}&
J^{+} + 2\bigl(\psi^{3}\psi^{+}\bigr) 
\nonumber\\
\cJ^{-} 
={}&
J^{-} -2\bigl(\psi^{3}\psi^{-}\bigr) \label{Jcal}
\\
\cJ^{3} 
={}&
J^{3} +\bigl(\psi^{-}\psi^{+}\bigr) \ . 
\nonumber
\end{align}
They then satisfy $\left[\cJ^{a}_{n},\psi^{b}_{r}\right]=0\,$, and define the same algebra as the $J^{a}_n$ with level $\kappa = k+2\,$. The Sugawara stress tensor and supercurrent are 
\begin{align}
T^{\mathfrak{sl}(2)}
={}&
 \frac{1}{2k}\left(\cJ^{+}\cJ^{-}+\cJ^{-}\cJ^{+} -2\, \cJ^{3}\cJ^{3}\right)-\frac{1}{2}\psi^{+}\partial \psi^{-}-\frac{1}{2}\psi^{-}\partial\psi^{+}+\psi^{3}\partial\psi^{3}
\label{Tdefsl} \\[4pt]
 G^{\mathfrak{sl}(2)}
 ={}&
 \frac{1}{\sqrt{k}}\left(\cJ^{+}\psi^{-}+\cJ^{-}\psi^{+} -2\, \cJ^{3}\psi^{3} - 2\psi^{+}\psi^{-}\psi^{3}\right)\ ,
 \label{Gdefsl}
\end{align}
where every composite operator in the above expressions is understood to be normal-ordered. They satisfy, in particular,
\be
\{G^{\mathfrak{sl}(2)}_r,\psi^a_s\} = J^a_{r+s} \ , \qquad [G^{\mathfrak{sl}(2)}_r,J^a_n] = - n \psi^a_{n+r} \ ,
\ee
\be
[L^{\mathfrak{sl}(2)}_n,\psi^a_s] = \bigl( -\tfrac{n}{2} - s \bigr) \, \psi^a_{n+s} \ , \qquad
[L^{\mathfrak{sl}(2)}_n,J^a_m] = - m J^a_{n+m} \ , 
\ee
as well as the relations of the ${\cal N}=1$ algebra
\begin{align}
\bigl[L^{\mathfrak{sl}(2)}_{m},L^{\mathfrak{sl}(2)}_{n}\bigr]
={}&
(m-n)L^{\mathfrak{sl}(2)}_{m+n} + \frac{c^{\mathfrak{sl}(2)}}{12}m\left(m^{2}-1\right)\delta_{m,-n}
\\
\bigl[L^{\mathfrak{sl}(2)}_{n},G^{\mathfrak{sl}(2)}_{r}\bigr] 
={}&
\left(\frac{n}{2}-r\right)G^{\mathfrak{sl}(2)}_{n+r}
\\
\bigl\{G^{\mathfrak{sl}(2)}_{r},G^{\mathfrak{sl}(2)}_{s}\bigr\}
={}&
2L^{\mathfrak{sl}(2)}_{r+s} + \frac{c^{\mathfrak{sl}(2)}}{3}\Bigl(r^{2}-\frac{1}{4}\Bigr)\delta_{r,-s}\ ,
\end{align}
where the central charge equals 
\be
c^{\mathfrak{sl}(2)} =  \frac{3(k+2)}{k} + \frac{3}{2} \ . 
\ee

\noindent For the two affine $\mathfrak{su}(2)$ algebras, we introduce the generators 
\begin{alignat}{5}
& \bigl[K^{(\pm)+}_{m},K^{(\pm)-}_{n}\bigr] 
={}
 2K^{(\pm)3}_{m+n} + k_{\pm}m\delta_{m,-n}
 &\quad &
 \bigl[K^{(\pm)3}_{m},K^{(\pm)\pm}_{n}\bigr] 
={}
\pm K^{(\pm)\pm}_{m+n}
 \nonumber\\
& \bigl[K^{(\pm)3}_{m},K^{(\pm)3}_{n}\bigr] 
={}
 \frac{k_{\pm}}{2}m\delta_{m,-n}
&\quad &
\bigl[K^{(\pm)\pm}_{m},\chi^{(\pm)3}_{r}\bigr]
 ={}
 \mp\chi^{(\pm) \pm}_{m+r} \nonumber \\
& \bigl[K^{(\pm)3}_{m},\chi^{(\pm)\pm}_{r}\bigr]
 ={}
 \pm\chi^{(\pm)\pm}_{m+r}
&\quad &
 \bigl[K^{(\pm)\pm}_{m},\chi^{(\pm)\mp}_{r}\bigr]
 ={}
 \pm 2\chi^{(\pm)3}_{m+r}
 \\
& \bigl\{\chi^{(\pm)+}_{r},\chi^{(\pm)-}_{s}\bigr\}
 ={}
 \delta_{r,-s}
 &\quad &
 \bigl\{\chi^{(\pm)3}_{r},\chi^{(\pm)3}_{s}\bigr\}
 ={}
 \frac{1}{2}\delta_{r,-s}\ .&
 \nonumber
\end{alignat}
Again, we can decouple the bosons from the fermions by defining the generators 
\begin{align}
\cK^{(\pm)+} 
={}&
K^{(\pm)+} - 2\bigl(\chi^{(\pm)3}\chi^{(\pm)+}\bigr) 
\nonumber\\
\cK^{(\pm)-} 
={}&
K^{(\pm)-} +2\bigl(\chi^{(\pm)3}\chi^{(\pm)-}\bigr) \label{Kcal}
\\
\cK^{(\pm)3} 
={}&
K^{(\pm)3} +\bigl(\chi^{(\pm)-}\chi^{(\pm)+}\bigr) \ . 
\nonumber
\end{align}
They then satisfy $[\cK^{(\pm)a}_{n},\chi^{(\pm)b}_{r}]=0\,$, and define the same algebra as the $K^{(\pm)a}_n$ with level 
\be
\kappa_{\pm} = k_{\pm}-2\ .
\ee
The Sugawara stress tensor and supercurrent are 
\begin{align}
T^{(\pm)}
={}&
 \frac{1}{2k_{\pm}}\left(\cK^{(\pm)+}\cK^{(\pm)-}+\cK^{(\pm)-}\cK^{(\pm)+} +2\, \cK^{(\pm)3}\cK^{(\pm)3} \right)\nonumber \\
 & \qquad  -\frac{1}{2}\chi^{(\pm)+}\partial \chi^{(\pm)-}-\frac{1}{2}\chi^{(\pm)-}\partial\chi^{(\pm)+}-\chi^{(\pm)3}\partial\chi^{(\pm)3}
\label{Tdefsu} \\[4pt]
 G^{(\pm)}
 ={}&
 \frac{1}{\sqrt{k_{\pm}}}\left(\cK^{(\pm)+}\chi^{(\pm)-}+\cK^{(\pm)-}\chi^{(\pm)+} +2\, \cK^{(\pm)3}\chi^{(\pm)3} + 2\chi^{(\pm)+}\chi^{(\pm)-}\chi^{(\pm)3}\right)\ ,
 \label{Gdefsu}
\end{align}
where again normal-ordering is understood. They satisfy, in particular,
\be
\{G^{(\pm)}_r,\chi^{(\pm)a}_s\} = K^{(\pm)a}_{r+s} \ , \qquad [G^{(\pm)}_r,K^{(\pm)a}_n] = - n \chi^{(\pm)a}_{n+r} \ ,
\ee
\be
[L^{(\pm)}_n,\chi^{(\pm)a}_s] = \bigl( -\tfrac{n}{2} - s \bigr) \, \chi^{(\pm)a}_{n+s} \ , \qquad
[L^{(\pm)}_n,K^{(\pm)a}_m] = - m K^{(\pm)a}_{n+m} \ , 
\ee
as well as the relations of the ${\cal N}=1$ algebra
\begin{align}
\bigl[L^{(\pm)}_{m},L^{(\pm)}_{n}\bigr]
={}&
(m-n)L^{(\pm)}_{m+n} + \frac{c^{(\pm)}}{12}m\left(m^{2}-1\right)\delta_{m,-n}
\\
\bigl[L^{(\pm)}_{n},G^{(\pm)}_{r}\bigr] 
={}&
\left(\frac{n}{2}-r\right)G^{(\pm)}_{n+r}
\\
\bigl\{G^{(\pm)}_{r},G^{(\pm)}_{s}\bigr\}
={}&
2L^{(\pm)}_{r+s} + \frac{c^{(\pm)}}{3}\Bigl(r^{2}-\frac{1}{4}\Bigr)\delta_{r,-s}\ ,
\end{align}
where the central charge equals now 
\be
c^{(\pm)} =  \frac{3(k_{\pm}-2)}{k_{\pm}} + \frac{3}{2} \ . 
\ee
Finally, we need the contribution from the $\mathfrak{u}(1)$ factor, which we shall describe in terms of a single boson $\partial \phi$ (with modes $\alpha$) and a single fermion $\lambda$ with commutation relations 
\be
{}[\alpha_m,\alpha_n] = m \delta_{m,-n} \ , \qquad \{ \lambda_r,\lambda_s\} = \frac{1}{2}\delta_{r,-s} \ .
\ee
The corresponding ${\cal N}=1$ generators are then 
\begin{eqnarray}
T^{\mathfrak{u}(1)} & = & \frac{1}{2} \partial \phi \partial\phi - \lambda \partial \lambda \nonumber \\
G^{\mathfrak{u}(1)} & = & \sqrt{2}\partial \phi \, \lambda \ . 
\end{eqnarray}
The total central charge of the whole worldsheet theory in eq.~(\ref{WZW}) is thus 
\be
c^{\rm total} = \frac{3 (k+2)}{k} +  \frac{3 (k_+-2)}{k_+} + \frac{3 (k_- -2)}{k_-} + 6 \ , 
\ee
and this equals $c^{\rm total}=15$ provided that 
\be
\frac{1}{k} = \frac{1}{k_+} + \frac{1}{k_-} \ ,
\ee
see eq.~(\ref{critical}).

\subsection{Modular invariance of the NS-R worldsheet theory} \label{app:NS-Rmod}

In this appendix we compute the partition function of the $\mathfrak{sl}(2,\mathds{R})$ part of the worldsheet theory of eq.~(\ref{wsspec}). Using \cite[eq.~(116)]{Maldacena:2000hw} we have that
\begin{equation}
Z(\tau,\theta)=e^{k\pi\frac{({\rm Im}\theta)^2}{{\rm Im}\tau}}\int_0^1d\alpha\int_{-\infty}^\infty ds\sum_{w\in\mathbb{Z}} |\chi_{\frac{1}{2}+is,\alpha,w}|^2\ , 
\end{equation}
where the $\mathfrak{sl}(2,\mathds{R})$ characters take the form
\begin{equation}
\chi_{\frac{1}{2}+is,\alpha,w}=2\pi \eta^{-3}e^{2\pi i\tau(\frac{s^2}{k-2}+\frac{kw^2}{4})}\sum_{m\in\mathbb{Z}}e^{2\pi im(\alpha+\frac{kw}{2})}\delta(\theta-w\tau-m)\ .
\end{equation}
Here $\theta$ is the chemical potential for $J^3_0$, and the exponential factor comes from the chiral anomaly, see \cite[Appendix B.3 \& B.4 and in particular eq.~(113)]{Maldacena:2000hw}. The integral over $\alpha$ gives rise to 
\begin{align}
\int_0^1d\alpha \, |\chi_{\frac{1}{2}+is,\alpha,w}|^2 = & 
\left|2\pi \eta^{-3}e^{2\pi i\tau(\frac{s^2}{k-2}+\frac{kw^2}{4})}\right|^2\, \sum_{m,n\in\mathbb{Z}}
\delta(\theta-w\tau-m)\, \delta(\bar\theta-w\bar\tau-n) \nonumber \\
& \qquad \quad \times \int_0^1d\alpha \, e^{2\pi i(m-n)(\alpha+\frac{kw}{2})} \ ,
\end{align}
and hence fixes $m=n$ so that the product of delta functions does not vanish. Thus, the partition function beomes
\begin{equation}
\begin{aligned}
\int_0^1d\alpha|\chi_{\frac{1}{2}+is,\alpha,w}|^2=& \left|2\pi \eta^{-3}e^{2\pi i\tau(\frac{s^2}{k-2}+\frac{kw^2}{4})}\right|^2\sum_{m\in\mathbb{Z}}\delta^{(2)}(\theta-w\tau-m)\,.
\end{aligned}
\end{equation}
Performing the integral over $s$ and the sum over $w$ gives then 
\begin{equation}
\begin{aligned}
Z(\tau;\theta)=e^{k\pi\frac{({\rm Im}\theta)^2}{{\rm Im}\tau}}\sum_w\frac{2\pi^2}{|\eta(\tau)|^6}\sqrt{\frac{k-2}{{\rm Im}\tau}}e^{-4\pi{\rm Im}\tau\frac{kw^2}{4}}\sum_{m\in\mathbb{Z}}\delta^{(2)}(\theta-w\tau-m)\,.
\end{aligned}
\end{equation}
The delta function restricts $w$ to be $w=\tfrac{{\rm Im}\theta}{{\rm Im}\tau}$, and hence, 
\begin{equation}
\begin{aligned}
Z(\theta;\tau)=&e^{k\pi\frac{({\rm Im}\theta)^2}{{\rm Im}\tau}}\frac{2\pi^2}{|\eta(\tau)|^6}\sqrt{\frac{k-2}{{\rm Im}\tau}}e^{-k\pi\frac{({\rm Im}\theta)^2}{{\rm Im}\tau}}\sum_{w,m\in\mathbb{Z}}\delta^{(2)}(\theta-w\tau-m)\\
=&\frac{2\pi^2}{|\eta(\tau)|^6}\sqrt{\frac{k-2}{{\rm Im}\tau}}\sum_{w,m\in\mathbb{Z}}\delta^{(2)}(\theta-w\tau-m)\,.
\end{aligned}
\end{equation}
This expression is then modular invariant. Indeed, the expression above is clearly $T$ invariant. To see the invariance under $S$, we note that the infinite sum gives rise to a factor $|\tau|^2$ upon $\tau\to-\tfrac{1}{\tau}$, $\theta\to\tfrac{\theta}{\tau}$. On the other hand, we have that ${\rm Im}(-\tfrac{1}{\tau})=\tfrac{{\rm Im}\tau}{|\tau|^2}$, and hence, the factors $|\tau|$ from the infinite sum and ${\rm Im}\tau$ cancel precisely with the $S$ transformation of $\eta$. 

\subsection{Spacetime supersymmetry}\label{appendix:spacetime susy}
In this appendix we review the construction of the spacetime supercharges for type IIB strings on $\rm AdS_3\times S^3\times\mathbb{T}^4$ with pure NS-NS flux following \cite{Giveon:1998ns}. Then we generalise the discussion to the background $\rm AdS_3\times S^3\times S^3\times S^1$. The discussion here applies for arbitrary values of the fluxes.

\subsubsection[\texorpdfstring{$\rm AdS_3\times S^3\times\mathbb{T}^4$}{AdS3xS3xT4}]{\boldmath \texorpdfstring{$\rm AdS_3\times S^3\times\mathbb{T}^4$}{AdS3xS3xT4}}\label{A.2.1}

In order to construct the spacetime supercharges it is convenient to bosonise the fermion bilinears in \eqref{eq:JT4} as follows
\begin{equation}
\begin{gathered}
i\partial H_1=\psi^+\psi^-,\quad i\partial H_2=\chi^+\chi^-,\quad i\partial H_3=2\psi^3\chi^3,\quad \\
i\partial H_4=-2i\lambda_1\lambda_2,\quad i\partial H_5=-2i\lambda_3\lambda_4\ , 
\end{gathered}
\end{equation}
where the free bosons $H_i$ then satisfy the OPE
\begin{equation}
H_i(z)H_j(w)\sim-\delta_{ij}\ln(z-w)\ .
\label{eq:free H_i OPEs}
\end{equation}
The spacetime supercharges in the $(-\tfrac{1}{2})$-picture can then be written as 
\begin{equation}
Q_\epsilon=\oint e^{-\frac{\phi}{2}}e^{\frac{i}{2}\epsilon_iH_i} \ , 
\label{eq:-1/2 picture of supercharge in T4}
\end{equation}
where the appropriate cocylce factors have been suppressed. Imposing mutual locality, only half of the supercharges survive, and we may take them to satisfy
\begin{equation}
\prod_{i=1}^5\epsilon_i=1\ .
\end{equation}
Next, we need to impose BRST invariance. The only nontrivial part comes from considering
\begin{equation}
\left[\oint dz(\gamma G^m)(z),Q_\epsilon\right]=\left[\oint dz(e^{\phi-\chi} G^m)(z),Q_\epsilon\right] = 0 \ .
\end{equation}
This requires that the OPE
\begin{equation}\label{fullOPE}
\bigl(e^{\phi-\chi} G^m\bigr)(z)\, \bigl(e^{-\frac{\phi}{2}}e^{\frac{i}{2}\epsilon_iH_i}\bigr)(w)
\end{equation}
is regular so that the above commutator vanishes. The singular terms in the OPE come from the trilinear fermion term in $G^m$, and they are schematically of the form 
\begin{equation}
\begin{aligned}
& G^m(z) \, (e^{\frac{i}{2}\epsilon_iH_i})(w) \sim 
\frac{-i}{\sqrt{k}}\left( 
\partial H_1(e^{iH_3}-e^{-iH_3})-\partial H_2(e^{iH_3}+e^{-iH_3}) \right)(z)e^{\frac{i}{2}\epsilon_iH_i}(w)\\
& \qquad \sim \frac{-\epsilon_1}{2\sqrt{k}\, (z-w)}\left( (z-w)^{\frac{\epsilon_3}{2}}+(z-w)^{-\frac{\epsilon_3}{2}}-\epsilon_1\epsilon_2(z-w)^{\frac{\epsilon_3}{2}}+\epsilon_1\epsilon_2(z-w)^{-\frac{\epsilon_3}{2}} \right)\ ,
\end{aligned}
\end{equation}
whereas the bosonised superghost OPE gives 
\be
e^{\phi}(z)\, e^{-\phi/2}(w)  \sim (z-w)^{\frac{1}{2}} \ . 
\ee
Thus in order to make the entire OPE in (\ref{fullOPE}) regular, we need to choose $\epsilon_3=-1$ if $\epsilon_1\epsilon_2=1$, and $\epsilon_3=1$ if $\epsilon_1\epsilon_2=-1$, i.e.\ 
\begin{equation}
\begin{aligned}
\epsilon_1\epsilon_2\epsilon_3=-1\,.
\end{aligned}
\end{equation}
Combining this with the mutual locality condition, the allowed supercharges are
\begin{equation}
Q_\epsilon=\oint e^{-\frac{\phi}{2}}\exp\left(\frac{i}{2}\left(\epsilon_1H_1+\epsilon_2H_2-\epsilon_1\epsilon_2H_3+\epsilon_4(H_4-H_5)\right)\right)\ .
\end{equation}
Finally, the `GSO projection' requires that physical operators must be mutually local with respect to these supercharges, and hence the physical operators must have integer charges with respect to the current
\begin{equation}
j_{\rm GSO}=\frac{1}{2}\partial\Bigl(\phi+i(H_1+H_2+H_3+H_4-H_5)\Bigr)\ .
\end{equation}
This explains why the fermion bilinear terms in $j_{\rm GSO}$ are precisely those that appear in \eqref{eq:JT4}, and it implies that the spectrum of $J$ in \eqref{eq:JT4} is indeed (half-)integer.

\subsubsection[\texorpdfstring{$\rm AdS_3\times S^3\times S^3\times S^1$}{AdS3xS3xS3xS1}]{\boldmath \texorpdfstring{$\rm AdS_3\times S^3\times S^3\times S^1$}{AdS3xS3xS3xS1}}\label{A.2.2}

For the case of $\rm AdS_3\times S^3\times S^3\times S^1$  the analysis is similar, except that not all supercharges will  turn out to be of the simple form of \eqref{eq:-1/2 picture of supercharge in T4}, see eq.~(\ref{eq:supercurrents with ``corrections''}) below. To start with we bosonise the fermions as 
\begin{equation}\label{ffield}
\begin{gathered}
i\partial H_1=\psi^+\psi^-,\quad i\partial H_2=\chi^{(+)+}\chi^{(+)-},\quad i\partial H_3=\chi^{(-)+}\chi^{(-)-},\quad \\
i\partial H_4=2\psi^3\left(\sqrt{\tfrac{k}{k_+}}\chi^{(+)3}+\sqrt{\tfrac{k}{k_-}}\chi^{(-)3}\right),\quad i\partial H_5=-2i\lambda\left(\sqrt{\tfrac{k}{k_-}}\chi^{(+)3}-\sqrt{\tfrac{k}{k_+}}\chi^{(-)3}\right)\ ,
\end{gathered}
\end{equation}
where the bosons again satisfy \eqref{eq:free H_i OPEs}. Let us first assume that the spacetime supercharges are of the form \eqref{eq:-1/2 picture of supercharge in T4}. Then mutual locality implies, as before, that we can only keep those which satisfy
\begin{equation}
\prod_{i=1}^5\epsilon_i=1\ .
\end{equation}
In order to study BRST invariance, we first solve (\ref{ffield}) for the individual fermions to find  
\begin{equation}
\begin{aligned}
\psi^3=&\frac{e^{iH_4}-e^{-iH_4}}{2}\ ,\\
\lambda=&\frac{e^{iH_5}+e^{-iH_5}}{2}\ ,\\
\chi^{(+)3}=&\frac{1}{\sqrt{k_++k_-}}\frac{\sqrt{k_-}(e^{iH_4}+e^{-iH_4})+\sqrt{k_+}(-ie^{iH_5}+ie^{-iH_5})}{2}\ ,\\
\chi^{(-)3}=&\frac{1}{\sqrt{k_++k_-}}\frac{\sqrt{k_+}(e^{iH_4}+e^{-iH_4})+\sqrt{k_-}(ie^{iH_5}-ie^{-iH_5})}{2}\ .
\end{aligned}
\end{equation}
The singular terms in the OPE (\ref{fullOPE}) are then schematically of the form 
\begin{equation}
\begin{aligned}
\frac{1}{(z-w)^{\frac{1}{2}}}&\Biggl[\Bigl(\frac{\epsilon_1}{\sqrt{k}}((z-w)^{\frac{\epsilon_4}{2}}+(z-w)^{-\frac{\epsilon_4}{2}}\Bigr)\\
&-\frac{\epsilon_2}{\sqrt{(k_++k_-)k_+}}\Bigl( \sqrt{k_-} \bigl((z-w)^{\frac{\epsilon_4}{2}}-(z-w)^{-\frac{\epsilon_4}{2}}\bigr)\\
&\hspace{4cm}-(-1)^{\frac{\epsilon_4}{2}}i\sqrt{k_+} \bigl((z-w)^{\frac{\epsilon_5}{2}}-(z-w)^{-\frac{\epsilon_5}{2}}\bigr) \Bigr)\\
&-\frac{\epsilon_3}{\sqrt{(k_++k_-)k_-}} \Bigl( \sqrt{k_+} \bigl((z-w)^{\frac{\epsilon_4}{2}}-(z-w)^{-\frac{\epsilon_4}{2}}\bigr)\\
&\hspace{4cm}+(-1)^{\frac{\epsilon_4}{2}}i\sqrt{k_-}\bigl((z-w)^{\frac{\epsilon_5}{2}}-(z-w)^{-\frac{\epsilon_5}{2}}\bigr) \Bigr)\Biggr]\ . \nonumber 
\end{aligned}
\end{equation}
Thus, the terms that involve $\epsilon_5$ imply that
\begin{equation}
\epsilon_2=\epsilon_3 \ ,
\label{eq:additional constraint in S3xS1}
\end{equation}
and the terms that involve $\epsilon_4$ imply that
\begin{equation}
\epsilon_1\epsilon_2\epsilon_4=-1\,,
\end{equation}
and thus from $\prod_{i=1}^5\epsilon_i=1$, we also have
\begin{equation}
\epsilon_5=-\epsilon_2\,.
\end{equation}
Hence, the spacetime supercurrents of the form \eqref{eq:-1/2 picture of supercharge in T4} are
\begin{equation}
Q_\epsilon=\oint e^{-\frac{\phi}{2}}e^{\frac{i}{2}(\epsilon_1H_1+\epsilon_2H_2+\epsilon_2H_3-\epsilon_1\epsilon_2H_4-\epsilon_2H_5)}\ ,
\label{eq:half of supercurrents in S3xS1}
\end{equation}
where $\epsilon_1,\epsilon_2\in\{\pm 1\}$. Again, mutual locality with $Q$ implies the integrality condition
\begin{equation}
\oint \frac{1}{2}\partial\Bigl(\phi+i(H_1+H_2+H_3+H_4 - H_5)\Bigr)\in\mathbb{Z}\,,
\end{equation}
and the $\mathfrak{u}(1)$ current in the (extended) $\mathcal{N}=2$ algebra should be given by \eqref{eq:JS3xS1}. 

However, because of the additional constraint \eqref{eq:additional constraint in S3xS1} that was not present for ${\rm AdS}_3\times {\rm S}^3 \times \mathbb{T}^4$, it would seem that there are now only $4$ spacetime supercharges as compared to the $8$ of the $\mathbb{T}^4$ case. The easiest way to generate the additional $4$ spacetime supercharges is to use the $\mathfrak{su}(2) \oplus \mathfrak{su}(2)$ $R$-symmetry. The associated DDF operators are of the form \cite{Giveon:1998ns}
\begin{equation}
\begin{aligned}
J^{(\pm)3}_0=&\oint \left(\chi^{(\pm)+}\chi^{(\pm)-}+\mathcal{K}^{(\pm)3}\right)=\oint \left(i\partial H_{2.5\mp0.5}+\mathcal{K}^{(\pm)3}\right)\ ,\\
J^{(\pm)\pm}_0=&\oint \left(\mathcal{K}^{(\pm)\pm}\mp2\chi^{(\pm)\pm}\chi^{(\pm)3}\right)\ .
\end{aligned}
\end{equation}
Following the discussion around eq.(3.15) of \cite{Giveon:1998ns}, this implies that the supercurrents \eqref{eq:half of supercurrents in S3xS1} carry the $\mathfrak{su}(2)_\pm$ labels 
\begin{equation}
\begin{aligned}
G^{++}_{-\frac{1}{2}}=&\oint e^{-\frac{\phi}{2}}e^{\frac{i}{2}(H_1+H_2+H_3-H_4-H_5)}\ ,\\
G^{--}_{-\frac{1}{2}}=&\oint e^{-\frac{\phi}{2}}e^{\frac{i}{2}(H_1-H_2-H_3+H_4+H_5)}\ ,\\
G^{++}_{\frac{1}{2}}=&\oint e^{-\frac{\phi}{2}}e^{\frac{i}{2}(-H_1+H_2+H_3+H_4-H_5)}\ ,\\
G^{--}_{\frac{1}{2}}=&\oint e^{-\frac{\phi}{2}}e^{\frac{i}{2}(-H_1-H_2-H_3-H_4+H_5)}\ . 
\end{aligned}
\end{equation}
The remaining supercharges can now be obtained by acting with $J^{(\pm)\pm}_0$, and we get (up to normalisation)
\begin{equation}
\begin{aligned}
G^{-+}_{-\frac{1}{2}}=&\frac{1}{\sqrt{k_++k_-}}\oint e^{-\frac{\phi}{2}}\Bigl( \sqrt{k_-}\, e^{\frac{i}{2}(H_1-H_2+H_3+H_4-H_5)}-\sqrt{k_+}\, e^{\frac{i}{2}(H_1-H_2+H_3-H_4+H_5)} \Bigr)\,,\\
G^{+-}_{-\frac{1}{2}}=&\frac{1}{\sqrt{k_++k_-}}\oint e^{-\frac{\phi}{2}}\Bigl( \sqrt{k_-}\, e^{\frac{i}{2}(H_1+H_2-H_3-H_4+H_5)}+\sqrt{k_+}\, e^{\frac{i}{2}(H_1+H_2-H_3+H_4-H_5)} \Bigr)\,,\\
G^{-+}_{\frac{1}{2}}=&\frac{1}{\sqrt{k_++k_-}}\oint e^{-\frac{\phi}{2}}\Bigl(\sqrt{k_-}\, e^{\frac{i}{2}(-H_1-H_2+H_3-H_4-H_5)} - \sqrt{k_+}\, e^{\frac{i}{2}(-H_1-H_2+H_3+H_4+H_5)}\Bigr)\,,\\
G^{+-}_{\frac{1}{2}}=&\frac{1}{\sqrt{k_++k_-}}\oint e^{-\frac{\phi}{2}}\Bigl( \sqrt{k_-}\, e^{\frac{i}{2}(-H_1+H_2-H_3+H_4+H_5)}+\sqrt{k_+}\, e^{\frac{i}{2}(-H_1+H_2-H_3-H_4-H_5)} \Bigr)\,.
\label{eq:supercurrents with ``corrections''}
\end{aligned}
\end{equation}
We have checked that these combinations then satisfy indeed BRST invariance.\footnote{For this calculation, the cocycle factors are critical.} Note that in the limit $k_\pm\rightarrow \infty$, only one of the two terms in each line survives, and we reproduce again the result for the $\mathbb{T}^4$ case.

\section{\boldmath The \texorpdfstring{$\mathfrak{d}(2,1|\alpha)$}{d(2,1|a)} algebra and the hybrid partition function}

The modes of the $\mathfrak{d}(2,1|\alpha)$ affine algebra satisfy the commutation and anti-commuta\-tion relations,
\begin{subequations}\label{eq:d21alpha}
\begin{align}
[J^3_m,J^3_n]&=-\tfrac{1}{2}km\delta_{m+n,0}\ ,   \label{eq:d21alpha commutation relations a}\\
[J^3_m,J^\pm_n]&=\pm J^\pm_{m+n}\ ,  \label{eq:d21alpha commutation relations b}\\
[J^+_m,J^-_n]&=km\delta_{m+n,0}-2J^3_{m+n}\ ,  \label{eq:d21alpha commutation relations c}\\
[K^{(\pm)3}_m,K^{(\pm)3}_n]&=\tfrac{1}{2}k^\pm m\delta_{m+n,0}\ ,  \label{eq:d21alpha commutation relations d}\\
[K^{(\pm)3}_m,K^{(\pm)\pm}_n]&=\pm K^{(\pm)\pm}_{m+n}\ ,  \label{eq:d21alpha commutation relations e}\\
[K^{(\pm)+}_m,K^{(\pm)-}_n]&=k^\pm m\delta_{m+n,0}+2K^{(\pm)3}_{m+n}\ , \label{eq:d21alpha commutation relations f}\\
[J^a_m,S^{\alpha\beta\gamma}_n]&=\tfrac{1}{2} \, c_a\, \tensor{(\sigma^a)}{^\alpha_\mu} S^{\mu\beta\gamma}_{m+n}\ ,  \label{eq:d21alpha commutation relations g}\\
[K^{(+)a}_m,S^{\alpha\beta\gamma}_n]&=\tfrac{1}{2}\tensor{(\sigma^a)}{^\beta_\nu} S^{\alpha\nu\gamma}_{m+n}\ ,  \label{eq:d21alpha commutation relations h}\\
[K^{(-)a}_m,S^{\alpha\beta\gamma}_n]&=\tfrac{1}{2}\tensor{(\sigma^a)}{^\gamma_\rho} S^{\alpha\beta\rho}_{m+n}\ , \label{eq:d21alpha commutation relations i}\\
 \{S^{\alpha\beta\gamma}_m,S^{\mu\nu\rho}_n\}&=km \varepsilon^{\alpha\mu}\varepsilon^{\beta\nu}\varepsilon^{\gamma\rho}\delta_{m+n,0}-\varepsilon^{\beta\nu}\varepsilon^{\gamma\rho} c_a\, \tensor{(\sigma_a)}{^{\alpha\mu}} J^a_{m+n}+\gamma\varepsilon^{\alpha\mu}\varepsilon^{\gamma\rho} \tensor{(\sigma_a)}{^{\beta\nu}} K^{(+)a}_{m+n}\nonumber\\
 &\qquad+(1-\gamma)\varepsilon^{\alpha\mu}\varepsilon^{\beta\nu} \tensor{(\sigma_a)}{^{\gamma\rho}} K^{(-)a}_{m+n}\ ,
 \label{eq:d21alpha commutation relations j}
\end{align}
\end{subequations}
where 
\be
\gamma = \frac{k_-}{k_-+ k_+} \ , 
\ee
while the `$\alpha$' parameter appearing in $\mathfrak{d}(2,1|\alpha)$ equals 
\be
\alpha = \frac{k_-}{k_+} = \frac{\gamma}{1-\gamma} \ . 
\ee
Thus for the case of primary interest, 
\be\label{alpha}
k_+ = k_- = 2 \ , \qquad \gamma = \frac{1}{2} \ , \qquad \alpha = 1 \ . 
\ee
Furthermore, $c_a=-1$ for $a=-$, and $c_a=1$ otherwise, and the non-zero $\sigma$ matrices are 
\be
\tensor{(\sigma_-)}{^{--}}=1\ , \qquad  \tensor{(\sigma_3)}{^{-+}}=1\ , \qquad \tensor{(\sigma_3)}{^{+-}}=1\ , \qquad  \tensor{(\sigma_+)}{^{++}}=-1\ .
\ee
Note that the index $a$ of $\sigma_a$ is raised (and lowered) by $\kappa^{ab}$, the Killing form, for which the nonzero components are
\begin{equation}
\kappa^{+-}=\kappa^{-+}=2\ ,\quad \kappa^{33}=1\ ,
\end{equation}
whereas the indices $\alpha,\beta$ are raised and lowered by $\epsilon^{\alpha\beta}$ and its inverse. We also clarify that, even though the indices are repeated in \eqref{eq:d21alpha commutation relations g}, there is no summation and the value of $a$ on the RHS is dictated by the value of $a$ on the LHS. Furtermore, the RHS of \eqref{eq:d21alpha commutation relations j} contains a term with 3 repeated indices. In this case, we mean that one should simply sum over all possible values of $a$. One can circumvent this abuse of notation by defining a new tensor 
\begin{equation}
\tensor{(\widetilde{\sigma}_a)}{^\alpha_\beta}:=c_a\tensor{(\sigma_a)}{^\alpha_\beta}\,,
\end{equation}
where, again, there is no summation on the RHS. Then eqs.\eqref{eq:d21alpha commutation relations g} and \eqref{eq:d21alpha commutation relations j} read
\begin{subequations}
\begin{align}
[J^a_m,S^{\alpha\beta\gamma}_n]&=\tfrac{1}{2}\tensor{(\widetilde{\sigma}_a)}{^\alpha_\mu} S^{\mu\beta\gamma}_{m+n}\ ,  \label{eq:d21alpha commutation relations g new}\\
 \{S^{\alpha\beta\gamma}_m,S^{\mu\nu\rho}_n\}&=km \varepsilon^{\alpha\mu}\varepsilon^{\beta\nu}\varepsilon^{\gamma\rho}\delta_{m+n,0}-\varepsilon^{\beta\nu}\varepsilon^{\gamma\rho} \tensor{(\widetilde{\sigma}_a)}{^{\alpha\mu}} J^a_{m+n}+\tfrac{1}{2}\varepsilon^{\alpha\mu}\varepsilon^{\gamma\rho} \tensor{(\sigma_a)}{^{\beta\nu}} K^{(+)a}_{m+n}\nonumber\\
 &\qquad+\tfrac{1}{2}\varepsilon^{\alpha\mu}\varepsilon^{\beta\nu} \tensor{(\sigma_a)}{^{\gamma\rho}} K^{(-)a}_{m+n}\,.
 \label{eq:d21alpha commutation relations j new}
\end{align}
\end{subequations}
The $\mathfrak{d}(2,1|\alpha)_k$ algebra has vanishing dual Coxeter number, and its central charge is therefore simply its superdimension, 
\be\label{cd}
c \bigl(\mathfrak{d}(2,1|\alpha)\bigr)= \hbox{sdim}\bigl(\mathfrak{d}(2,1|\alpha)\bigr) = 9 - 8 = 1 \ ,
\ee
irrespective of the level. 

\subsection{Modular invariance }\label{app:hybrid}

In the remainder of this appendix we show that the spectrum given in eq.~(\ref{hybrid_p}) is indeed modular invariant. We begin by calculating the coset character from the free field realisation.

\subsubsection{The coset character}\label{app:hybridchar}

Let us first determine the coset character without spectral flow. We start with the character of the free field realisation 
\begin{equation}\label{freefield}
\chi_{\rm free}(t,z_\pm,\mu;\tau) = {\rm Tr}_{\rm free}\left( q^{L_0-\frac{c}{24}}e^{2\pi iz_+K^{(+)3}_0}e^{2\pi iz_-K^{(-)3}_0}e^{2\pi i\mu Z_0} \right)\ .
\end{equation}
Following the analysis in Appendix C of \cite{Eberhardt:2018ouy},  the character of the continuous representation of two pairs of symplectic
bosons $(\mathcal{E}_{\frac{\alpha+is}{2}},\mathcal{E}_{\frac{\alpha-is}{2}})$ is given by
\begin{equation}
\begin{aligned}\label{sympbos}
&\frac{1}{\eta^4(\tau)}\sum_{m\in\frac{\mathbb{Z}}{2}+\frac{\alpha+is}{2},\ n\in\frac{\mathbb{Z}}{2}+\frac{\alpha-is}{2}}x^{m+n}e^{2\pi i(-2\mu)(m-n-\frac{1}{2})}\\
&=\frac{1}{\eta^4(\tau)}\Bigl( \sum_{r\in\mathbb{Z}+\alpha,\ u\in\mathbb{Z}+is}+\sum_{r\in\mathbb{Z}+\alpha+\frac{1}{2},\ u\in\mathbb{Z}+\frac{1}{2}+is} \Bigr) \, x^re^{2\pi i(-2\mu)(u-\frac{1}{2})}\ ,
\end{aligned}
\end{equation}
whereas the character for the 8 fermions is
\begin{equation}
\begin{aligned}\label{fermions}
\frac{\vartheta_2(\frac{z_++z_-+2\mu}{2};\tau)\vartheta_2(\frac{z_++z_--2\mu}{2};\tau)\vartheta_2(\frac{z_+-z_--2\mu}{2};\tau)\vartheta_2(\frac{z_+-z_-+2\mu}{2};\tau)}{\eta^4(\tau)}\ .
\end{aligned}
\end{equation}
The free field character in eq.~(\ref{freefield}) is then the product of these two expressions. Next we need to decompose this free field character in terms of the $\mathfrak{u}(1)$ coset representations, i.e.\ we need to write 
\begin{equation}
\chi_{\rm free}(t,z_\pm,\mu;\tau)=\sum_{s,\alpha} \chi_{{\cal R}(s,\alpha)}(t,z_\pm,\mu;\tau) \, \frac{q^{- s^2}}{\eta(\tau)} \ , 
\end{equation}
where the states ${\cal R}(s,\alpha)$ arise from the states with $Z_0 = - 2 i s$. 
In order to extract the part of $\chi_{\rm free}$ with this $Z_0$ charge, we first write 
\begin{equation}
\vartheta_2\Bigl(\frac{z+t}{2};\tau\Bigr)\vartheta_2\Bigl(\frac{z-t}{2};\tau\Bigr)=\vartheta_2(z;2\tau)\vartheta_3(t,2\tau)+\vartheta_3(z;2\tau)\vartheta_2(t,2\tau) \ ,
\label{eq:product of theta2 identity}
\end{equation}
and thus rewrite the product of the four theta functions in eq.~(\ref{fermions}) as
\begin{equation}
\begin{aligned}\label{productform}
&\left[ \vartheta_2\Bigl(\frac{z_++z_-}{2};2\tau\Bigr)\vartheta_3 (2\mu;2\tau)+\vartheta_3\Bigl(\frac{z_++z_-}{2};2\tau\Bigr)\vartheta_2(2\mu;2\tau) \right]\\
\times&\left[ \vartheta_2\Bigl(\frac{z_+-z_-}{2};2\tau\Bigr)\vartheta_3(2\mu;2\tau)+\vartheta_3\Bigl(\frac{z_+-z_-}{2};2\tau\Bigr)\vartheta_2(2\mu;2\tau) \right]\ .
\end{aligned}
\end{equation}
Let us begin by considering the term $\vartheta_3(2\mu;2\tau)\vartheta_3(2\mu;2\tau)$, which we can expand as 
\begin{equation}
\vartheta_3(2\mu;2\tau)\vartheta_3(2\mu;2\tau)=\sum_{m,n\in\mathbb{Z}}q^{m^2+n^2}e^{2\pi i(m+n)(2\mu)}\,.
\end{equation}
Combining this with the sum over $u$ from \eqref{sympbos}, we see that we obtain altogether 
\begin{equation}
\sum_{u\in\mathbb{Z}+\frac{\epsilon}{2}+is} \sum_{m,n\in\mathbb{Z}}q^{m^2+n^2}e^{2\pi i(m+n)(2\mu)}e^{2\pi i(-2\mu)(u-\frac{1}{2})}\ ,
\end{equation}
where $\epsilon=0,1$, depending on whether we consider the first or second sum in \eqref{sympbos}. The term corresponding to $Z_0 = - 2is$ now arises provided that $u=m+n+\tfrac{1}{2}+is$, i.e.\ it picks out one term from the second sum with $\epsilon=1$ for each pair $(m,n)$. Finally, the sum over $m,n$ gives the theta functions 
\begin{equation}
\sum_{m,n\in\mathbb{Z}}q^{m^2+n^2}=\vartheta_3^2(2\tau)\ .
\end{equation}
Proceeding similarly for the other terms in the product (\ref{productform}), the coefficient of $e^{2\pi i \mu (-2 i s)}$ in the free field character is thus (up to an overall factor of $\eta^{-8}$)
\begin{equation}
\begin{aligned}
&\sum_{r\in\mathbb{Z}+\alpha+\frac{1}{2}}x^r\Bigl( 
\vartheta_2(\frac{z_++z_-}{2};2\tau)\vartheta_2(\frac{z_+-z_-}{2};2\tau)\vartheta_3^2(2\tau) \\
& \qquad \qquad +\vartheta_3(\frac{z_++z_-}{2};2\tau)\vartheta_3(\frac{z_+-z_-}{2};2\tau)\vartheta_2^2(2\tau) \Bigr)\\[4pt]
&\quad +\sum_{r\in\mathbb{Z}+\alpha}x^r\Bigl( 
\vartheta_2(\frac{z_++z_-}{2};2\tau)\vartheta_3(\frac{z_+-z_-}{2};2\tau) \\
& \qquad \qquad +\vartheta_3(\frac{z_++z_-}{2};2\tau)\vartheta_2(\frac{z_+-z_-}{2};2\tau) \Bigr)\vartheta_3(2\tau)\vartheta_2(2\tau)
\end{aligned}
\end{equation}
\begin{equation}
\begin{aligned}
&=\sum_{r\in\mathbb{Z}+\alpha+\frac{1}{2}}x^r \Bigl( 
\vartheta_2(\frac{z_++z_-}{2};2\tau)\vartheta_2(\frac{z_+-z_-}{2};2\tau)\vartheta_3^2(t;2\tau) \\
& \qquad \qquad + \vartheta_3(\frac{z_++z_-}{2};2\tau)\vartheta_3(\frac{z_+-z_-}{2};2\tau)\vartheta_2^2(t;2\tau) \Bigr)\\[4pt]
&\quad +\sum_{r\in\mathbb{Z}+\alpha+\frac{1}{2}}x^r\Bigl( 
\vartheta_2(\frac{z_++z_-}{2};2\tau)\vartheta_3(\frac{z_+-z_-}{2};2\tau)\\
& \qquad \qquad +\vartheta_3(\frac{z_++z_-}{2};2\tau)\vartheta_2(\frac{z_+-z_-}{2};2\tau) \Bigr)\vartheta_3(t;2\tau)\vartheta_2(t;2\tau)\ , 
\end{aligned}
\end{equation}
where in the second step we have shifted the sum over $r$ and absorbed some of the $x$-factors into the  $\vartheta_i(2\tau)$ functions (that now also depend on $t$). Note that this expression is independent of the charge $Z_0=-2is$. 

Recombining these terms using again \eqref{eq:product of theta2 identity} and dividing by the denominator character $q^{- s^2}/\eta(\tau)$, we thus finally obtain 
\newpage 

\begin{align}\label{C.11}
& \chi_{{\cal R}(s,\alpha)}(t,z_\pm,\mu=0;\tau) \nonumber \\
& \qquad 
=q^{s^2}\sum_{r\in\mathbb{Z}+\alpha+\frac{1}{2}}x^r\frac{\vartheta_2(\frac{z_++z_-+t}{2};\tau)\vartheta_2(\frac{z_++z_--t}{2};\tau)\vartheta_2(\frac{z_+-z_--t}{2};\tau)\vartheta_2(\frac{z_+-z_-+t}{2};\tau)}{\eta^7(\tau)}\ .
\end{align}
We have also confirmed, at least for the first few terms, that all of these states appear in the irreducible $\mathfrak{d}(2,1|\alpha)_1$ representation whose highest weight states are described by eq.~(\ref{eq: long repn in d(2,1|alpha)}); thus eq.~(\ref{C.11}) is (probably) the character of this irreducible representation.\footnote{It is clear that the coset contains $\mathfrak{d}(2,1|\alpha)_1$, but, a priori, it could describe a bigger algebra.}
\smallskip

Applying spectral flow, see eq.~(\ref{specflow}), is now straightforward, and we find for the character of the spectrally flowed representation 
\begin{align}
\chi^{(w)}_{{\cal R}(s,\alpha)}(t,z_\pm;\tau) & = q^{s^2+\frac{w^2}{4}} x^{\frac{w}{2}}\, \sum_{r\in\mathbb{Z}+\alpha+\frac{1}{2}}x^r q^{-rw} \\
& \qquad \times 
\frac{\vartheta_2(\frac{z_++z_-+t}{2};\tau)\vartheta_2(\frac{z_++z_--t}{2};\tau)\vartheta_2(\frac{z_+-z_--t}{2};\tau)\vartheta_2(\frac{z_+-z_-+t}{2};\tau)}{\eta^7(\tau)}\ .  \nonumber 
\end{align}
By shifting the parameter $r$ by $\frac{w}{2}$, this can be rewritten as
\begin{align}
 \chi^{(w)}_{{\cal R}(s,\alpha)}(t,z_\pm;\tau) & = q^{s^2+\frac{3 w^2}{4}} \, \sum_{r\in\mathbb{Z}+\alpha+\frac{w+1}{2}}x^rq^{-rw} \\
& \qquad \times 
 \frac{\vartheta_2(\frac{z_++z_-+t}{2};\tau)\vartheta_2(\frac{z_++z_--t}{2};\tau)\vartheta_2(\frac{z_+-z_--t}{2};\tau)\vartheta_2(\frac{z_+-z_-+t}{2};\tau)}{\eta^7(\tau)}\ ,\nonumber\\
 & = q^{s^2+\frac{3 w^2}{4}} \, \sum_{m\in\mathbb{Z}} e^{2\pi i m( \alpha+\frac{w+1}{2} )}\, \delta(t-w\tau-m) \label{eq(3.37)}\\
& \qquad \times 
 \frac{\vartheta_2(\frac{z_++z_-+t}{2};\tau)\vartheta_2(\frac{z_++z_--t}{2};\tau)\vartheta_2(\frac{z_+-z_--t}{2};\tau)\vartheta_2(\frac{z_+-z_-+t}{2};\tau)}{\eta^7(\tau)}\,.\nonumber 
\end{align}

\subsubsection{Modular invariance}

With these preparations we can now study the modular invariance of the full partition function of eq.~(\ref{hybrid_p}). It is convenient to first strip off the $s$-dependence from the coset character by removing the factor of 
\begin{equation}
\vartheta_s(\tau):=\frac{q^{s^2}}{\eta(\tau)} \ , 
\end{equation}
whose $S$-transformation is simply 
\begin{equation}
\vartheta_s\bigl(-\tfrac{1}{\tau}\bigr)=\sqrt{2}\int_{\mathbb{R}}dt\, e^{4\pi its}\vartheta_t(\tau) = 
\int_{\mathbb{R}}dt\, S_{s,t}\vartheta_t(\tau)\ , \qquad S_{s,t}=\sqrt{2}e^{4\pi its}\ .
\end{equation}
If we denote the remainder by 
\be
\hat{\chi}^{(w)}_{{\cal R}(\alpha)}(t,z_\pm;\tau) \equiv \frac{\eta(\tau)}{q^{s^2}}\, \chi^{(w)}_{{\cal R}(s,\alpha)}(t,z_\pm;\tau)  \ ,
\ee
its $S$-transformation $(t,z_\pm,\tau)\to(\tfrac{t}{\tau},\tfrac{z_\pm}{\tau},-\tfrac{1}{\tau})$ equals 
\begin{equation}
\begin{aligned}
&e^{\frac{\pi i}{\tau}(\frac{t^2}{2}-z_+^2-z_-^2)} \hat\chi^{(w)}_{{\cal R}(\alpha)}\bigl(\tfrac{t}{\tau},\tfrac{z_\pm}{\tau};-\tfrac{1}{\tau}\bigr)\\[4pt]
&\hspace{1cm}=i\, {\rm sgn}({\rm Re}(\tau))\sum_{w'\in\mathbb{Z}}q^{\frac{3w'^2}{4}}e^{2\pi iw'\alpha}(-1)^{w'}\delta(t-w'\tau+w)\\
&\hspace{1cm}\quad \times\frac{\vartheta_2(\frac{z_++z_-+t}{2};\tau)\vartheta_2(\frac{z_++z_--t}{2};\tau)\vartheta_2(\frac{z_+-z_--t}{2};\tau)\vartheta_2(\frac{z_+-z_-+t}{2};\tau)}{\eta^6(\tau)} \ , \label{3.38}
\end{aligned}
\end{equation}
which we can rewrite as 
\be
 \sum_{w'\in\mathbb{Z}} \int_0^1 d\alpha' \, \hat{S}_{(w,\alpha),(w',\alpha')} \, \hat{\chi}^{(w')}_{{\cal R}(\alpha')}(t,z_\pm;\tau) \ , 
 \label{eq:S transformation with S matrix}
\ee
with 
\begin{equation}
\hat{S}_{(w,\alpha),(w',\alpha')}=i\, {\rm sgn}({\rm Re}(\tau))\, e^{2\pi i (w'\alpha+w\alpha'+\frac{ww'}{2}+\frac{w+w'}{2})} \ . 
\end{equation}
To see this, let us start with \eqref{eq:S transformation with S matrix}, and first do the $\alpha'$ integral,  
using \eqref{eq(3.37)}, which gives the Kronecker delta $\delta_{m,-w}$. This then collapses the sum over $m$ to $m=-w$,  and we find 
\begin{equation}
\begin{aligned}
&\sum_{w',m\in\mathbb{Z}} \int_0^1 d\alpha'\, e^{2\pi i (w'\alpha+w\alpha'+\frac{ww'}{2}+\frac{w+w'}{2})}
q^{\frac{3w'^2}{4}}\, e^{2\pi i m ( \alpha'+\frac{w'+1}{2} )}\, \delta(t-w'\tau-m) \\
&=\sum_{w',m\in\mathbb{Z}} q^{\frac{3w'^2}{4}}\delta_{m,-w}\, e^{2\pi i (w'\alpha+\frac{ww'}{2}+\frac{w+w'}{2})}\, e^{2\pi i m ( \frac{w'+1}{2} )}\, \delta(t-w'\tau-m) \\
&=\sum_{w'\in\mathbb{Z}} q^{\frac{3w'^2}{4}}\, e^{2\pi i (w'\alpha+\frac{w'}{2})}  \delta(t-w'\tau+w)\ ,
\end{aligned}
\end{equation}
which therefore reproduces the sum in eq.~(\ref{3.38}).

It is then straightforward to see that the full partition function is modular invariant. Indeed, both the ${\rm S}^1$ part and the terms proportional to $s$ are modular invariant by themselves; for the compact ${\rm S}^1$ factor this is well known, and for the latter this is simply a consequence of the fact that 
\begin{equation}
\begin{aligned}
\int_{\mathbb{R}}ds\,|\vartheta_s(-1/\tau)|^2=&\int_{\mathbb{R}}ds\, dt\, dt' \, S_{s,t}\bar S_{s,t'}\vartheta_t(\tau)\overline{\vartheta_{t'}(\tau)}\,,\\=&2\int_{\mathbb{R}}dt\, dt'\, \int_{\mathbb{R}}ds\, e^{4\pi is(t-t')}\vartheta_t(\tau)\overline{\vartheta_{t'}(\tau)} \,\ ,\\
=&\int_{\mathbb{R}}dt\, dt'\, \delta(t-t')\vartheta_t(\tau)\overline{\vartheta_{t'}(\tau)} = 
\int_{\mathbb{R}}dt|\vartheta_t(\tau)|^2 \ . 
\end{aligned}
\end{equation}
This leaves us with showing that the remainder, i.e.\ 
\be
Z_{\rm full-2 bosons}= \sum_{w\in\mathbb{Z}} \int_0^1d\alpha\,  | \hat\chi^{(w)}_{{\cal R}(\alpha)}|^2  \ , 
\ee
is modular invariant, which follows from 
\begin{equation}
\begin{aligned}
& \sum_{w\in\mathbb{Z}} \int_0^1d\alpha\, |\hat{\chi}^{(w)}_{{\cal R}(\alpha)}(-1/\tau)|^2 \\
& \qquad = \int_0^1d\alpha \, d\alpha'\, d\alpha''\!\!\! \!\!\sum_{w,w',w''\in\mathbb{Z}}\!\!\! S_{(w,\alpha),(w',\alpha
')}\overline{S}_{(w,\alpha),(w'',\alpha
'')}\hat{\chi}^{(w')}_{{\cal R}(\alpha')}\overline{\hat{\chi}^{(w'')}_{{\cal R}(\alpha'')}} \\
& \qquad = \int_0^1 d\alpha'd\alpha''\sum_{w',w''\in\mathbb{Z}}\delta_{w',w''}\sum_{m\in\mathbb{Z}}\delta(\alpha'-\alpha''-m)\hat{\chi}^{(w')}_{{\cal R}(\alpha')}\overline{\hat{\chi}^{(w'')}_{{\cal R}(\alpha'')}} \\
& \qquad = \int_0^1 d\alpha'\sum_{w'\in\mathbb{Z}}|\hat{\chi}^{(w')}_{{\cal R}(\alpha')}(\tau)|^2\ , 
\end{aligned}
\end{equation}
where, in going to the second line, we have used that
\begin{equation}
\begin{aligned}
& \int_0^1d\alpha\sum_{w\in\mathbb{Z}}S_{(w,\alpha),(w',\alpha ')}\overline{S}_{(w,\alpha),(w'',\alpha '')} \\ 
& \qquad \qquad =\sum_{w\in\mathbb{Z}}\int_0^1\, d\alpha\, e^{2\pi i ((w'-w'')\alpha+w(\alpha'-\alpha'')+\frac{w(w'-w'')}{2}+\frac{(w'-w'')}{2})} \\
& \qquad \qquad =\sum_{w\in\mathbb{Z}}\delta_{w',w''}\, e^{2\pi i (w(\alpha'-\alpha'')+\frac{w(w'-w'')}{2}+\frac{(w'-w'')}{2})} \\
& \qquad \qquad = \delta_{w',w''}\, \sum_{w\in\mathbb{Z}} \, e^{2\pi iw(\alpha'-\alpha'')} = \delta_{w',w''}\, \sum_{m\in\mathbb{Z}}\delta(\alpha'-\alpha''-m)\ .
\end{aligned}
\end{equation}
Thus the full partition function is modular invariant as claimed.

\bibliographystyle{JHEP}

\end{document}